# Neural Encoding and Decoding with Deep Learning for Dynamic Natural Vision

Haiguang Wen[2,3], Junxing Shi[2,3], Yizhen Zhang[2,3], Kun-Han Lu[2,3], Jiayue Cao[1,3], Zhongming Liu[*1,2,3]

**Convolutional neural network (CNN) driven by image recognition has been shown to be able to explain cortical responses to static pictures at ventral-stream areas. Here, we further showed that such CNN could reliably predict and decode functional magnetic resonance imaging data from humans watching natural movies, despite its lack of any mechanism to account for temporal dynamics or feedback processing. Using separate data, encoding and decoding models were developed and evaluated for describing the bi-directional relationships between the CNN and the brain. Through the encoding models, the CNN-predicted areas covered not only the ventral stream, but also the dorsal stream, albeit to a lesser degree; single-voxel response was visualized as the specific pixel pattern that drove the response, revealing the distinct representation of individual cortical location; cortical activation was synthesized from natural images with high-throughput to map category representation, contrast, and selectivity. Through the decoding models, fMRI signals were directly decoded to estimate the feature representations in both visual and semantic spaces, for direct visual reconstruction and semantic categorization, respectively. These results corroborate, generalize, and extend previous findings, and highlight the value of using deep learning, as an all-in-one model of the visual cortex, to understand and decode natural vision.**

Neural encoding | brain decoding | deep learning | natural vision

## Introduction

For centuries, philosophers and scientists have been trying to speculate, observe, understand, and decipher the workings of the brain that enables humans to perceive and explore visual surroundings. Here, we ask how the brain represents dynamic visual information from the outside world, and whether brain activity can be directly decoded to reconstruct and categorize what a person is seeing. These questions, concerning neural encoding and decoding (Naselaris et al., 2011), have been mostly addressed with static or artificial stimuli (Kamitani and Tong, 2005; Haynes and Rees, 2006). Such strategies are, however, too narrowly focused to reveal the computation underlying natural vision. What is needed is an alternative strategy that embraces the complexity of vision to uncover and decode the visual representations of distributed cortical activity.

Despite its diversity and complexity, the visual world is composed of a large number of visual features (Zeiler and Fergus, 2014; LeCun et al., 2015; Russ and Leopold, 2015). These features span many levels of abstraction, such as orientation and color in the low level, shapes and textures in the middle levels, and objects and actions in the high level. To date, deep learning provides the most comprehensive computational models to encode and extract hierarchically organized features from arbitrary natural pictures or videos (LeCun et al., 2015). Computer-vision systems based on such models have emulated or even surpassed human performance in image recognition and segmentation (Krizhevsky et al., 2012; Russakovsky et al., 2015; He et al., 2015). In particular, deep convolutional neural networks (CNN) are built and trained with similar organizational and coding principles as the feedforward visual cortical network (DiCarlo et al., 2012; Yamins and DiCarlo, 2016). Recent studies have shown that the CNN could partially explain the brain's responses to (Yamins et al., 2014; Güçlü and van Gerven, 2015a; Eickenberg et al., 2016) and representations of (Khaligh-Razavi and Kriegeskorte, 2014; Cichy et al., 2016) natural picture stimuli. However, it remains unclear whether and to what extent the CNN may explain and decode brain responses to natural video stimuli. Although dynamic natural vision involves feedforward, recurrent, and feedback connections (Callaway, 2004), the CNN only models feedforward processing and operates on instantaneous input, without any account for recurrent or feedback network interactions (Polack and Contreras, 2012; Bastos et al., 2012).

To address these questions, we acquired 11.5 hours of fMRI data from each of three human subjects watching 972 different video clips, including diverse scenes and actions. This dataset was independent of, and had a larger sample size and broader coverage than, those in prior studies (Güçlü and van Gerven, 2015a; Eickenberg et al., 2016; Khaligh-Razavi and Kriegeskorte, 2014; Yamins et al., 2014; Güçlü and van Gerven, 2015a;

*Correspondence: zmliu@purdue.edu
[1]Weldon School of Biomedical Engineering, [2]School of Electrical and Computer Engineering, [3]Purdue Institute for Integrative Neuroscience, Purdue University, West Lafayette, Indiana, 47906, USA.

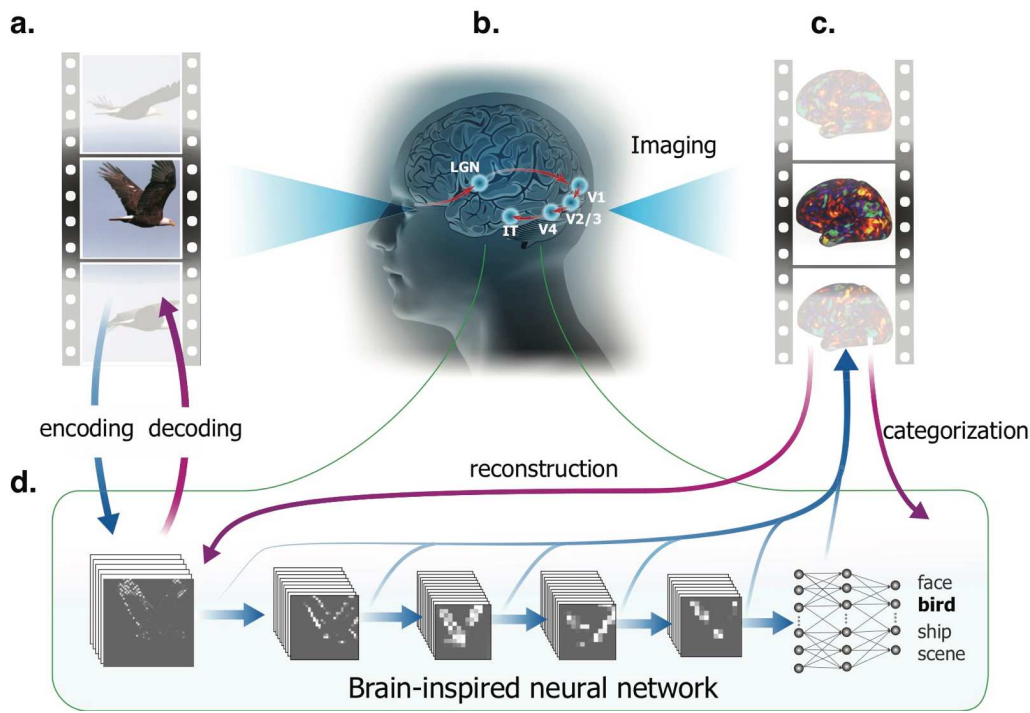

**Figure. 1. Neural encoding and decoding through a deep learning model**. When a person is seeing a film **(a)**, information is processed through a cascade of cortical areas **(b)**, generating fMRI activity patterns **(c)**. A deep convolutional neural network is used here to model cortical visual processing **(d)**. This model transforms every movie frame into multiple layers of features, ranging from orientations and colors in the visual space (the 1$^{st}$ layer) to object categories in the semantic space (the 8$^{th}$ layer). For encoding, this network serves to model the nonlinear relationship between the movie stimuli and the response at each cortical location. For decoding, cortical responses are combined across locations to estimate the feature outputs from the 1$^{st}$ and 7$^{th}$ layer. The former is deconvolved to reconstruct every movie frame, and the latter is classified into semantic categories.

Eickenberg et al., 2016; Cichy et al., 2016). This allowed us to confirm, generalize, and extend the use of the CNN in predicting and decoding cortical activity along both ventral and dorsal streams in a dynamic viewing condition. Specifically, we trained and tested encoding and decoding models, with distinct data, for describing the relationships between the brain and the CNN, implemented by (Krizhevsky et al., 2012). With the CNN, the encoding models were used to predict and visualize fMRI responses at individual cortical voxels given the movie stimuli; the decoding models were used to reconstruct and categorize the visual stimuli based on fMRI activity, as shown in Fig. 1. The major findings were

1) a CNN driven for image recognition explained significant variance of fMRI responses to complex movie stimuli for nearly the entire visual cortex including its ventral and dorsal streams, albeit to a lesser degree for the dorsal stream;

2) the CNN-based voxel-wise encoding models visualized different single-voxel representations, and revealed category representation and selectivity;

3) the CNN supported direct visual reconstruction of natural movies, highlighting foreground objects with blurry details and missing colors;

4) the CNN also supported direct semantic categorization, utilizing the semantic space embedded in the CNN.

## Materials and Methods

### Subjects & Experiments

Three healthy volunteers (female, age: 22-25; normal vision) participated in the study, with informed written consent obtained from every subject according to the research protocol approved by the Institutional Review Board at Purdue University. Each subject was instructed to watch a series of natural color video clips (20.3°!20.3 °) while fixating at a central fixation cross (0.8°!0.8 °). In total, 374 video clips (continuous with a frame rate of 30 frames per second) were included in a 2.4-hour training movie, randomly split into 18 8-min segments; 598 different video clips were included in a 40-min testing movie, randomly split into five 8-min segments. The video clips in the testing movie were different from those in the training movie. All video clips were chosen from Videoblocks (https://www.videoblocks.com) and YouTube (https://www.youtube.com) to be diverse yet representative of real-life visual experiences. For example, individual video clips showed people in action, moving animals, nature scenes, outdoor or indoor scenes etc. Each subject watched the training movie twice and the testing movie ten times through experiments in different days. Each experiment included multiple sessions of 8min and 24s long. During each session, an 8-min single movie segment was presented; before the movie presentation, the first movie frame was displayed as a static picture for 12 seconds; after the movie, the last movie frame was also displayed as a static picture for 12 seconds. The order of the movie segments was randomized and counter-balanced. Using Psychophysics Toolbox 3 (http://psychtoolbox.org), the visual stimuli were delivered through a goggle system (NordicNeuroLab NNL Visual System) with 800!600 display resolution.

**Data Acquisition and Preprocessing**

$T_1$ and $T_2$-weighted MRI and fMRI data were acquired in a 3 tesla MRI system (Signa HDx, General Electric Healthcare, Milwaukee) with a 16-channel receive-only phase-array surface coil (NOVA Medical, Wilmington). The fMRI data were acquired at 3.5 mm isotropic spatial resolution and 2 s temporal resolution by using a single-shot, gradient-recalled echo-planar imaging sequence (38 interleaved axial slices with 3.5mm thickness and 3.5!3.5mm$^2$ in-plane resolution, TR=2000ms, TE=35ms, flip angle=78°, field of view=22!22cm$^2$). The fMRI data were preprocessed and then transformed onto the individual subjects' cortical surfaces, which were co-registered across subjects onto a cortical surface template based on their patterns of myelin density and cortical folding. The preprocessing and registration were accomplished with high accuracy by using the processing pipeline for the Human Connectome Project (Glasser et al., 2013). When training and testing the encoding and decoding models (as described later), the cortical fMRI signals were averaged over multiple repetitions: two repetitions for the training movie, and 10 repetitions for the testing movie. The two repetitions of the training movie allowed us to evaluate intra-subject reproducibility in the fMRI signal as a way to map the regions "activated" by natural movie stimuli (see **Mapping cortical activations with natural movie stimuli**). The ten repetitions of the testing movie allowed us to obtain the movie-evoked responses with high signal to noise ratios (SNR), as spontaneous activity or noise unrelated to visual stimuli were effectively removed by averaging over this relatively large number of repetitions. The ten repetitions of the testing movie also allowed us to estimate the upper bound (or "noise ceiling"), by which an encoding model could predict the fMRI signal during the testing movie. Although more repetitions of the training movie would also help to increase the SNR of the training data, it was not done because the training movie was too long to repeat by the same times as the testing movie.

**Convolutional Neural Network (CNN)**

We used a deep CNN (a specific implementation referred as the "AlexNet") to extract hierarchical visual features from the movie stimuli. The model had been pre-trained to achieve the best-performing object recognition in Large Scale Visual Recognition Challenge 2012 (Krizhevsky et al., 2012). Briefly, this CNN included eight layers of computational units stacked into a hierarchical architecture: the first five were convolutional layers, and the last three layers were fully connected for image-object classification (Supplementary Fig. 1). The image input was fed into the first layer; the output from one layer served as the input to its next layer. Each convolutional layer contained a large number of units and a set of filters (or kernels) that extracted filtered outputs from all locations from its input through a rectified linear function. Layer 1 through 5 consisted of 96, 256, 384, 384, and 256 kernels, respectively. Max-pooling was implemented between layer 1 and layer 2, between layer 2 and layer 3, and between layer 5 and layer 6. For classification, layer 6 and 7 were fully connected networks; layer 8 used a softmax function to output a vector of probabilities, by which an input image was classified into individual categories. The numbers of units in layer 6 to 8 were 4096, 4096, and 1000, respectively.

Note that the 2$^{nd}$ highest layer in the CNN (i.e. the 7$^{th}$ layer) effectively defined a semantic space to support the categorization at the output layer. In other words, the semantic information about the input image was represented by a (4096-dimensional) vector in this semantic space. In the original AlexNet, this semantic space was used to classify ~1.3 million natural pictures into 1,000 fine-grained categories (Krizhevsky et al., 2012). Thus, it was generalizable and inclusive enough to also represent the semantics in our training and testing movies, and to support more coarsely defined categorization. Indeed, new classifiers could be built for image classification into new categories based on the generic representations in this same semantic space, as shown elsewhere for transfer learning (Razavian et al., 2014).

Many of the 1,000 categories in the original AlexNet were not readily applicable to our training or testing movies. Thus, we reduced the number of categories to 15 for mapping categorical representations and decoding object categories from fMRI. The new categories were coarser and labeled as *indoor*, *outdoor*, *people*, *face*, *bird*, *insect*, *water animal*, *land animal*, *flower*,

*fruit*, *natural scene*, *car*, *airplane*, *ship*, and *exercise*. These categories covered the common content in both the training and testing movies. With the redefined output layer, we trained a new softmax classifier for the CNN (i.e. between the 7$^{th}$ layer and the output layer), but kept all lower layers unchanged. We used ~20,500 human-labeled images to train the classifier while testing it with a different set of ~3,500 labeled images. The training and testing images were all randomly and evenly sampled from the aforementioned 15 categories in ImageNet, followed by visual inspection to replace mislabeled images.

In the softmax classifier (a multinomial logistic regression model), the input was the semantic representation, $y$, from the 7$^{th}$ layer in the CNN, and the output was the normalized probabilities, $q$, by which the image was classified into individual categories. The softmax classifier was trained by using the mini-batch gradient descend to minimize the Kullback-Leibler (KL) divergence from the predicted probability, $q$, to the ground truth, $p$, in which the element corresponding to the labeled category was set to one and others were zeros. The KL divergence indicated the amount of information lost when the predicted probability, $q$, was used to approximate $p$. The predicted probability was expressed as $q = \frac{\exp(y\mathbf{W}+\mathbf{b})}{\sum \exp(y\mathbf{W}+\mathbf{b})}$, parameterized with $\mathbf{W}$ and $\mathbf{b}$. The objective function that was minimized for training the classifier was expressed as below.

$$D_{KL}(\mathbf{p} \| \mathbf{q}) = H(\mathbf{p},\mathbf{q}) - H(\mathbf{p}) = -\langle \mathbf{p}, \log \mathbf{q} \rangle + \langle \mathbf{p}, \log \mathbf{p} \rangle \quad (1)$$

where $H(\mathbf{p})$ was the entropy of $\mathbf{p}$, and $H(\mathbf{p},\mathbf{q})$ was the cross-entropy of $\mathbf{p}$ and $\mathbf{q}$, and $\langle \cdot \rangle$ stands for inner product. The objective function was minimized with L2-norm regularization whose parameter was determined by cross-validation. 3075 validation images (15% of the training images) were uniformly and randomly selected from each of the 15 categories. When training the model, the batch size was 128 samples per batch, the learning rate was initially 10$^{-3}$ reduced by 10$^{-6}$ every iteration. After training with 100 epochs, the classifier achieved a top-1 error of 13.16% with the images in the testing set.

Once trained, the CNN could be used for feature extraction and image recognition by a simple feedforward pass of an input image. Specifically, passing a natural image into the CNN resulted in an activation value at each unit. Passing every frame of a movie resulted in an activation time series from each unit, representing the fluctuating representation of a specific feature in the movie. Within a single layer, the units that shared the same kernel collectively output a feature map given every movie frame. Herein we refer to the output from each layer as the output of the rectified linear function before max-pooling (if any).

**Deconvolutional neural network (De-CNN)**

While the CNN implemented a series of cascaded "bottom-up" transformations that extracted nonlinear features from an input image, we also used the De-CNN to approximately reverse the operations in the CNN, for a series of "top-down" projections as described in detail elsewhere (Zeiler and Fergus, 2014). Specifically, the outputs of one or multiple units could be unpooled, rectified, and filtered onto its lower layer, until reaching the input pixel space. The unpooling step was only applied to the layers that implemented max-pooling in the CNN. Since the max-pooling was non-invertible, the unpooling was an approximation while the locations of the maxima within each pooling region were recorded and used as a set of switch variables. Rectification was performed as point-wise rectified linear thresholding by setting the negative units to 0. The filtering step was done by applying the transposed version of the kernels in the CNN to the rectified activations from the immediate higher layer, to approximate the inversion of the bottom-up filtering. In the De-CNN, rectification and filtering were independent of the input, whereas the unpooling step was dependent on the input. Through the De-CNN, the feature representations at a specific layer could yield a reconstruction of the input image (Zeiler and Fergus, 2014). This was utilized for reconstructing the visual input based on the 1$^{st}$-layer feature representations estimated from fMRI data (see details in **Reconstructing natural movie stimuli** in **Methods**). Such reconstruction is unbiased by the input image, since the De-CNN did not perform unpooling from the 1$^{st}$ layer to the pixel space.

**Mapping cortical activations with natural movie stimuli**

Each segment of the training movie was presented twice to each subject. This allowed us to map cortical locations activated by natural movie stimuli, by computing the intra-subject reproducibility in voxel time series (Hasson et al., 2004; Lu et al., 2016). For each voxel and each segment of the training movie, the intra-subject reproducibility was computed as the correlation of the fMRI signal when the subject watched the same movie segment for the first time and for the second time. After converting the correlation coefficients to z scores by using the Fisher's z-transformation, the voxel-wise z scores were averaged across all 18 segments of the training movie. Statistical significance was evaluated by using one-sample t-test (p<0.01, DOF=17, Bonferroni correction for the number of cortical voxels), revealing the cortical regions activated by the training movie. Then, the intra-subject reproducibility maps were averaged across the three subjects. The averaged activation map was used to create a cortical mask that covered all significantly activated locations. To be more generalizable to other subjects or stimuli, we slightly expanded the mask. The final mask contained 10,214 voxels in the visual cortex, approximately 17.2% of the whole cortical surface.

**Bivariate analysis to relate CNN units to brain voxels**

We compared the outputs of CNN units to the fMRI signals at cortical voxels during the training movie, by evaluating the correlation between every unit and every voxel. Before this bivariate correlation analysis, the single unit activity in the CNN was log-transformed and convolved with a canonical hemodynamic response function (HRF) with the positive peak at 4s. Such preprocessing was to account for the difference in distribution, timing, and sampling between the unit activity and the fMRI signal. The unit activity was non-negative and sparse; after log-transformation (i.e. $\log(y + 0.01)$ where y indicated the unit activity), it followed a distribution similar to that of the fMRI signal. The HRF accounted for the temporal delay and smoothing due to neurovascular coupling. Here, we preferred a pre-defined HRF to a model estimated from the fMRI data itself. While the latter was data-driven and used in previous studies (Nishimoto et al., 2011; Güçlü and van Gerven, 2015b), it might cause over-fitting. A pre-defined HRF was suited for more conservative estimation of the bivariate (unit-to-voxel) relationships. Lastly, the HRF-convolved unit activity was down-sampled to match the sampling rate of fMRI. With such preprocessing, the bivariate correlation analysis was used to map the retinotopic, hierarchical, and categorical representations during natural movie stimuli, as described subsequently.

**Retinotopic mapping.** In the first layer of the CNN, individual units extracted features (e.g. orientation-specific edge) from different local (11-by-11 pixels) patches in the input image. We computed the correlation between the fMRI signal at each cortical location and the activation time series of every unit in the first layer of the CNN during the training movie. For a given cortical location, such correlations formed a 3-D array: two dimensions corresponding to the horizontal and vertical coordinates in the visual field, and the third dimension corresponding to 96 different local features (see Fig. 7c). As such, this array represented the simultaneous tuning of the fMRI response at each voxel by retinotopy, orientation, color, contrast, spatial frequency etc. We reduced the 3-D correlation array into a 2-D correlation matrix by taking the maximal correlation across different visual features. As such, the resulting correlation matrix depended only on retinotopy, and revealed the population receptive field (pRF) of the given voxel. The pRF center was determined as the centroid of the top 20 locations with the highest correlation values, and its polar angle and eccentricity were further measured with respect to the central fixation point. Repeating this procedure for every cortical location gave rise to the putative retinotopic representation of the visual cortex. We compared this retinotopic representation obtained with natural visual stimuli to the visual-field maps obtained with the standard retinotopic mapping as previously reported elsewhere (Abdollahi et al., 2014).

**Hierarchical mapping.** The feedforward visual processing passes through multiple cascaded stages in both the CNN and the visual cortex. In line with previous studies (Khaligh-Razavi and Kriegeskorte, 2014; Yamins et al., 2014; Güçlü and van Gerven, 2015a,b; Cichy et al., 2016; Kubilius et al., 2016; Horikawa and Kamitani, 2017; Eickenberg et al., 2016), we explored the correspondence between individual layers in the CNN and individual cortical regions underlying different stages of visual processing. For this purpose, we computed the correlations between the fMRI signal at each cortical location and the activation time series from each layer in the CNN, and extracted the maximal correlation. We interpreted this maximal correlation as a measure of how well a cortical location corresponded to a layer in the CNN. For each cortical location, we identified the best corresponding layer and assigned its layer index to this location; the assigned layer index indicated the processing stage this location belonged to. The cortical distribution of the layer-index assignment provided a map of the feedforward hierarchical organization of the visual system.

**Mapping representations of object categories.** To explore the correspondence between the high-level visual areas and the object categories encoded by the output layer of the CNN, we examined the cortical fMRI correlates to the 15 categories output from the CNN. Here, we initially focused on the "face" because face recognition was known to involve specific visual areas, such as the fusiform face area (FFA) (Kanwisher et al., 1997; Johnson, 2005). We computed the correlation between the activation time series of the face-labeled unit (the unit labeled as "face" in the output layer of the CNN) and the fMRI signal at every cortical location, in response to each segment of the training movie. The correlation was then averaged across segments and subjects. The significance of the average correlation was assessed using a block permutation test (Adolf et al., 2014) in consideration of the auto-correlation in the fMRI signal. Specifically, the time series was divided into 50-sec blocks of adjacent 25 volumes (TR=2s). The block size was chosen to be long enough to account for the autocorrelation of fMRI and to ensure a sufficient number of permutations to generate the null distribution. During each permutation step, the "face" time series underwent a random shift (i.e. removing a random number of samples from the beginning and adding them to the end) and then the time-shifted signal was divided into blocks, and per-muted by blocks. For a total of 100,000 times of permutations, the correlations between the fMRI signal and the permuted "face" time series was calculated. This procedure resulted in a realistic null distribution, against which the p value of the correlation (without permutation) was calculated with Bonferroni correction by the number of voxels. The significantly correlated voxels (p<0.01) were displaced to reveal cortical regions responsible for the visual processing of human faces. The same strategy was also applied to the mapping of other categories.

**Voxel-wise encoding models**

Furthermore, we attempted to establish the CNN-based predictive models for the fMRI response to natural movie stimuli. Such models were defined separately for each voxel, namely voxel-wise encoding models (Naselaris et al., 2011), through which the voxel response was predicted from a linear combination of the feature representations of the input movie. Conceptually similar encoding models were previously explored with low-level visual features (Kay et al., 2008; Nishimoto et al., 2011) or high-level semantic features (Huth et al., 2012, 2016a), and more recently with hierarchical features extracted by the CNN from static pictures (Güçlü and van Gerven, 2015a; Eickenberg et al., 2016). Here, we extended these prior studies to focus on natural movie stimuli while using principal component analysis (PCA) to reduce the huge dimension of the feature space attained with the CNN.

Specifically, PCA was applied to the feature representations obtained from each layer of the CNN during the training movie. Principal components were retained to keep 99% of the variance while spanning a much lower-dimensional feature space, in which the representations followed a similar distribution as did the fMRI signal. This dimension reduction mitigated the potential risk of overfitting with limited training data. In the reduce feature space, the feature time series were readily comparable with the fMRI signal without additional non-linear (log) transformation.

Mathematically, let $\mathbf{Y}_o^l$ be the output from all units in layer $l$ of the CNN; it is an $m$-by-$p$ matrix ($m$ is the number of video frames in the training movie, and $p$ is the number of units). The time series extracted by each unit was standardized (i.e. remove the mean and normalize the variance). Let $\mathbf{B}^l$ be the principal basis of $\mathbf{Y}_o^l$; it is a $p$-by-$q$ matrix ($q$ is the number of components). Converting the feature representations from the unit-wise space to the component-wise space is expressed as below.

$$\mathbf{Y}_n^l = \mathbf{Y}_o^l \mathbf{B}^l \qquad (2)$$

where $\mathbf{Y}_n^l$ is the transformed feature representations in the dimension-reduced feature space spanned by unitary columns in the matrix, $\mathbf{B}^l$. The transpose of $\mathbf{B}^l$ also defined the transformation back to the original space.

Following the dimension reduction, the feature time series, $\mathbf{Y}_n^l$, were convolved with a HRF, and then down-sampled to match the sampling rate of fMRI. Hereafter, $\mathbf{Y}^l$ stands for the feature time series for layer $l$ after convolution and down-sampling. These feature time series were used to predict the fMRI signal at each voxel through a linear regression model, elaborated as below.

Given a voxel $v$, the voxel response $\boldsymbol{x}_v$ was modeled as a linear combination of the feature time series, $\mathbf{Y}^l$, from the $l$-th layer in the CNN, as expressed in Eq. (3).

$$\boldsymbol{x}_v = \mathbf{Y}^l \boldsymbol{w}_v^l + b_v^l + \boldsymbol{\varepsilon} \qquad (3)$$

where, $\boldsymbol{w}_v^l$ is a $q$-by-1 vector of the regression coefficients; $b_v^l$ is the bias term; $\boldsymbol{\varepsilon}$ is the error unexplained by the model. Least-squares estimation with L2-norm regularization, as Eq. (4), was used to estimate the regression coefficients based on the data during the training movie.

$$f(\boldsymbol{w}_v^l) = \|\boldsymbol{x}_v - \mathbf{Y}^l \boldsymbol{w}_v^l - b_v^l\|_2^2 + \lambda \|\boldsymbol{w}_v^l\|_2^2 \qquad (4)$$

Here, the L2 regularization was used to prevent the model from overfitting limited training data. The regularization parameter $\lambda$ and the layer index $l$ were both optimized through a nine-fold cross-validation. Briefly, the training data were equally split into nine subsets: eight for the model estimation, one for the model validation. The validation was repeated nine times such that each subset was used once for validation. The parameters ($\lambda$, $l$) were chosen to maximize the cross-validation accuracy. With the optimized parameters, we refitted the model using the entire training samples to yield the final estimation of the voxel-wise encoding model. The final encoding model set up a computational pathway from the visual input to the evoked fMRI response at each voxel via its most predictive layer in the CNN.

After training the encoding model, we tested the model's accuracy in predicting the fMRI response to all five segments of the testing movie, for which the model was not trained. For each voxel, the prediction accuracy was measured as the correlation between the measured fMRI response and the response predicted by the voxel-specific encoding model, averaged across the segments of the testing movie. The significance of the correlation was assessed using a block permutation test (Adolf et al., 2014), while considering the auto-correlation in the fMRI signal, similarly as the significance test for the unit-to-voxel correlation (see **Mapping representations of object categories** in **Methods**). Briefly, the predicted fMRI signal was randomly block-permuted in time for 100,000 times to generate an empirical null distribution, against which the prediction accuracy was evaluated for significance (p<0.001, Bonferroni correction by the number of voxels). The prediction accuracy was also evaluated for regions of interest (ROIs) defined with multi-modal cortical parcellation (Glasser et al., 2016). For the ROI analysis, the voxel-wise prediction accuracy was averaged within each ROI. The prediction accuracy was evaluated for each subject, and then compared and averaged across subjects.

The prediction accuracy was compared with an upper bound by which the fMRI signal was explainable by the visual stimuli, given the presence of noise or ongoing activity unrelated to the stimuli. This upper bound, defining the explainable variance for each voxel, depended on the signal to noise ratio of the evoked fMRI response. It was measured voxel by voxel based on the fMRI signals observed during repeated presentations of the testing movie. Specifically, 10 repetitions of the testing movie were divided by half. This two-half partition defined an (ideal) control model: the signal averaged within the first half was used to predict the signal averaged within the second half. Their correlation, as the upper bound of the prediction accuracy, was compared with the prediction accuracy obtained with the voxel-wise encoding model in predicting the same testing data. The difference between their prediction accuracies (z-score) was assessed by paired t-test (p<0.01) across all possible two-half partitions and all testing movie segments. For those significant voxels, we then calculated the percentage of the explainable variance that was not explained by the encoding model. Specifically, let $V_c$ be the potentially explainable variance; let $V_e$ be the variance explained by the encoding model; so, $(V_c - V_e)/V_c$ measures the degree by which the encoding falls short in explaining the stimulus-evoked response (Wu et al., 2006).

**Predicting cortical responses to images and categories**

After testing their ability to predict cortical responses to unseen stimuli, we further used the encoding models to predict voxel-wise cortical responses to arbitrary pictures. Specifically, 15,000 images were uniformly and randomly sampled from 15 categories in ImageNet (i.e. *face, people, exercise, bird, land-animal, water-animal, insect, flower, fruit, car, airplane, ship, natural scene, outdoor, indoor*). None of these sampled images were used to train the CNN, or included in the training or testing movies. For each sampled image, the response at each voxel was predicted by using the voxel-specific encoding model. The voxel's responses to individual images formed a response profile, indicative of its selectivity to single images.

To quantify how a voxel selectively responded to images from a given category (e.g. face), the voxel's response profile was sorted in a descending order of its response to every image. Since each category contained 1,000 exemplars, the percentage of the top-1000 images belonging to one category was calculated as an index of the voxel's categorical selectivity. This selectivity index was tested for significance using a binomial test against a null hypothesis that the top 1,000 images were uniformly random across individual categories. This analysis was tested specifically for voxels in the fusiform face area (FFA).

For each voxel, its categorical representation was obtained by averaging single-image responses within categories. The representational difference between inanimate vs. animate categories was assessed, with former including *flower, fruit, car, airplane, ship, natural scene, outdoor, indoor*, and the latter including *face, people, exercise, bird, land-animal, water-animal, insect*. The significance of this difference was assessed with two-sample t-test with Bonferroni correction by the number of voxels.

**Visualizing single-voxel representations**

The voxel-wise encoding models set up a computational path to relate any visual input to the evoked fMRI response at each voxel. It inspired and allowed us to reveal which part of the visual input specifically accounted for the response at each voxel, or to visualize the voxel's representation of the input. Note that the visualization was targeted to each voxel, as opposed to a layer or unit in the CNN, as in (Güçlü and van Gerven, 2015a). This distinction was important because voxels with activity predictable by the same layer in the CNN, may bear highly or entirely different representations.

Let us denote the visual input as $\mathbf{I}$. The response $\mathbf{x}_v$ at a voxel $v$ was modeled as $\mathbf{x}_v = \mathrm{E}_v(\mathbf{I})$ ($\mathrm{E}_v$ is the voxel's encoding model). The voxel's visualized representation was an optimal gradient pattern given the visual input $\mathbf{I}$ that reflected the pixel-wise influence in driving the voxel's response. This optimization included two steps, combining the visualization methods based on masking (Zhou et al., 2014; Li, 2016) and gradient (Baehrens et al., 2010; Hansen et al., 2011; Simonyan et al., 2013; Springenberg et al., 2014).

Firstly, the algorithm searched for an optimal binary mask, $\mathbf{M}_o$, such that the masked visual input gave rise to the maximal response at the target voxel, as Eq. (5).

$$\mathbf{M}_o = \arg\max_{\mathbf{M}}\{\mathrm{E}_v(\mathbf{I} \circ \mathbf{M})\} \qquad (5)$$

where the mask was a 2-D matrix with the same width and height as the visual input $\mathbf{I}$, and $\circ$ stands for the Hadamard product, meaning that the same masking was applied to the red, green, and blue channels respectively. Since the encoding model was highly nonlinear and not convex, random optimization (Matyas, 1965) was used. A binary continuous mask (i.e. the pixel weights were either 1 or 0) was randomly and iteratively generated.

For each iteration, a random pixel pattern was generated with each pixel's intensity sampled from a normal distribution; this random pattern was spatially smoothed with a Gaussian spatial-smoothing kernel (three times of the kernel size of 1st layer CNN units); the smoothed pattern was thresholded by setting one fourth pixels to 1 and others 0. Then, the model-predicted response was computed given the masked input. The iteration was stopped when the maximal model-predicted response (over all iterations) converged or reach 100 iterations. The optimal mask was the one with the maximal response across iterations.

After the mask was optimized, the input from the masked region, $\mathbf{I}_o = \mathbf{I} \circ \mathbf{M}_o$, was supplied to the voxel-wise encoding model. The gradient of the model's output was computed with respect to the intensity at every pixel in the masked input, as expressed by Eq. (6). This gradient pattern described the relative influence of every pixel in driving the voxel response. Only positive gradients, which indicated the amount of influence in increasing the voxel response, were back-propagated and kept, as in (Springenberg et al., 2014).

$$\mathbf{G}_v(\mathbf{I}_o) = \nabla \mathrm{E}_v(\mathbf{I})|_{\mathbf{I}=\mathbf{I}_o} \quad (6)$$

For the visualization to be more robust, the above two steps were repeated 100 times. The weighted average of the visualizations across all repeats was obtained with the weight proportional to the response given the masked input for each repeat (indexed with $i$), as Eq. (7). Consequently, the averaged gradient pattern was taken as the visualized representation of the visual input at the given voxel.

$$\mathbf{G}_v(\mathbf{I}_o) = \frac{1}{100} \sum_{i=1}^{100} \mathbf{G}_v^i(\mathbf{I}_o) \mathrm{E}_v^i(\mathbf{I}_o) \quad (7)$$

This visualization method was applied to the fMRI signals during one segment of the testing movie. To explore and compare the visualized representations at different cortical locations, example voxels were chosen from several cortical regions across different levels, including V2, V4, MT, LO, FFA and PPA. Within each of these regions, we chose the voxel with the highest average prediction accuracy during the other four segments of the testing movie. The single-voxel representations were visualized only at time points where peak responses occurred at one or multiple of the selected voxels.

**Reconstructing natural movie stimuli**

Opposite to voxel-wise encoding models that related visual input to fMRI signals, decoding models transformed fMRI signals to visual and semantic representations. The former was used to reconstruct the visual input, and the latter was used to uncover its semantics.

For the visual reconstruction, multivariate linear regression models were defined to take as input the fMRI signals from all voxels in the visual cortex, and to output the representation of every feature encoded by the 1st layer in the CNN. As such, the decoding models were feature-wise and multivariate. For each feature, the decoding model had multiple inputs and multiple outputs (i.e. representations of the given feature from all spatial locations in the visual input), and the times of fMRI acquisition defined the samples for the model's input and output. Eq. (8) describes the decoding model for each of 96 different visual features.

$$\mathbf{Y} = \mathbf{XW} + \boldsymbol{\varepsilon} \quad (8)$$

Here, $\mathbf{X}$ stands for the observed fMRI signals within the visual cortex. It is an $m$-by-$(k+1)$ matrix, where $m$ is the number of time points, $k$ is the number of voxels; the last column of $\mathbf{X}$ is a constant vector with all elements equal to 1. $\mathbf{Y}$ stands for the log-transformed time-varying feature map. It is an $m$-by-$p$ matrix, where $m$ is the number of time points, and $p$ is the number of units that encode the same local image feature (i.e. the convolutional kernel). $\mathbf{W}$ stands for the unknown weights, by which the fMRI signals are combined across voxels to predict the feature map. It is an $(k+1)$-by-$p$ matrix with the last row being the bias component. $\boldsymbol{\varepsilon}$ is the error term.

To estimate the model, we optimized $\mathbf{W}$ to minimize the objective function below.

$$f(\mathbf{W}) = \|\mathbf{Y} - \mathbf{XW}\|_2^2 + \lambda \|\mathbf{W}\|_1^1 \quad (9)$$

where the first term is the sum of squares of the errors; the second term is the L1 regularization on $\mathbf{W}$ except for the bias component; $\lambda$ is the hyperparameter balancing these two terms. Here, L1 regularization was used rather than L2 regularization, since the former favored sparsity as each visual feature in the 1st CNN layer was expected to be coded by a small set of voxels in the visual cortex (Kay et al., 2008; Olshausen and Field, 1996).

The model estimation was based on the data collected with the training movie. $\lambda$ was determined by 20-fold cross-validation, similar to the procedures used for training the encoding models. For training, we used stochastic gradient descent optimization with the batch size of 100 samples, i.e. only 100 fMRI volumes were utilized in each iteration of training. To address the overfitting problem, dropout technique (Srivastava et al., 2014) was used by randomly dropping 30% of voxels in every iteration, i.e. setting the dropped voxels to zeros. Dropout regularization was used to mitigate the co-linearity among voxels and counteract L1 regularization to avoid over-sparse weights. For the cross-validation, we evaluated for each of the 96 features, the validation accuracy defined as the correlation between the fMRI-estimated feature map and the CNN-extracted feature map. After sorting the different features in a descending order of the validation accuracy, we identified those features with relatively low cross-validation accuracy ($r < 0.24$), and excluded them when reconstructing the testing movie.

To test the trained decoding model, we applied it to the fMRI signals observed during one of the testing movies, according to Eq. (8) without the error term. To evaluate the performance of the decoding model, the fMRI-estimated feature maps were correlated with those extracted from the CNN given the testing movie. The correlation coefficient, averaged across different features, was used as a measure of the accuracy for visual reconstruction. To test the statistical significance of the reconstruction accuracy, a block permutation test was performed. Briefly, the estimated feature maps were randomly block-permuted in time (Adolf et al., 2014) for 100,000 times to generate an empirical null distribution, against which the estimation accuracy was evaluated for significance ($p<0.01$), similar to the aforementioned statistical test for the voxel-wise encoding model.

To further reconstruct the testing movie from the fMRI-estimated feature maps, the feature maps were individually converted to the input pixel space using the De-CNN, and then were summed to generate the reconstruction of each movie frame. It is worth noting that the De-CNN did not perform unpooling from the 1$^{st}$ layer to the pixel space; so, the reconstruction was unbiased by the input, making the model generalizable for reconstruction of any unknown visual input. As a proof of concept, the visual inputs could be successfully reconstructed through De-CNN given the accurate (noiseless) feature maps (Supplementary Fig. S13).

**Semantic categorization**

In addition to visual reconstruction, the fMRI measurements were also decoded to deduce the semantics of each movie frame at the fMRI sampling times. The decoding model for semantic categorization included two steps: 1) converting the fMRI signals to the semantic representation of the visual input in a generalizable semantic space, 2) converting the estimated semantic representation to the probabilities by which the visual input belonged to pre-defined and human-labeled categories.

In the first step, the semantic space was spanned by the outputs from the 7$^{th}$ CNN layer, which directly supported the image classification at the output layer. This semantic space was generalizable to not only novel images, but also novel categories which the CNN was not trained for (Razavian et al., 2014). As defined in Eq. (10), the decoding model used the fMRI signals to estimate the semantic representation, denoted as $\mathbf{Y}_s$ ($m$-by-$q$ matrix, where $q$ is the dimension of the dimension-reduced semantic space and $m$ is the number of time points).

$$\mathbf{Y}_s = \mathbf{X}\mathbf{W}_s + \boldsymbol{\varepsilon} \quad (10)$$

where $\mathbf{X}$ stands for the observed fMRI signals within the visual cortex, and $\mathbf{W}_s$ was the regression coefficients, and $\boldsymbol{\varepsilon}$ was the error term. To train this decoding model, we used the data during the training movie and applied L2-regularization. The estimated dimension-reduced representation was then transformed back to the original space. The regularization parameter and $q$ were determined by 9-fold cross validation based on the correlation between estimated representation and the ground truth.

In the second step, the semantic representation estimated in the first step was converted to a vector of normalized probabilities over categories. This step utilized the softmax classifier established when retraining the CNN for image classification into 15 labeled categories (see **Convolutional Neural Network** in **Methods**).

After estimating the decoding model with the training movie, we applied it to the data during one of the testing movies. It resulted in the decoded categorization probability for individual frames in the testing movie sampled every 2 seconds. The top 5 categories with the highest probabilities were identified, and their textual labels were displayed as the semantic descriptions of the reconstructed testing movie.

To evaluate the categorization accuracy, we used top-1 through top-3 prediction accuracies. Specifically, for any given movie frame, we ranked the object categories in a descending order of the fMRI-estimated probabilities. If the true category was the top 1 of the ranked categories, it was considered to be top-1 accurate. If the true category was in the top 2 of the ranked categories, it was considered to be top-2 accurate, so on and so forth. The percentage of the frames that were top-1/top-2/top-3 accurate was calculated to quantify the overall categorization accuracy, for which the significance was evaluated by a binomial test against the null hypothesis that the categorization accuracy was equivalent to the chance level given random guesses. Note that the ground-truth categories for the testing movie was manually labeled by human observers, instead of the CNN's categorization of the testing movie.

**Cross-subject encoding and decoding**

To explore the feasibility of establishing encoding and decoding models generalizable to different subjects, we first evaluated the inter-subject reproducibility of the fMRI voxel response to the same movie stimuli. For each segment of the training movie, we calculated for each voxel the correlation of the fMRI signals between different subjects. The voxel-wise correlation coefficients were z-transformed and then averaged across all segments of the training movie. We assessed the significance of the reproducibility against zeros by using one-sample t-test with the degree of freedom as the total number of movie segments minus 1 (DOF=17, Bonferroni correction for the number of voxels, and p<0.01).

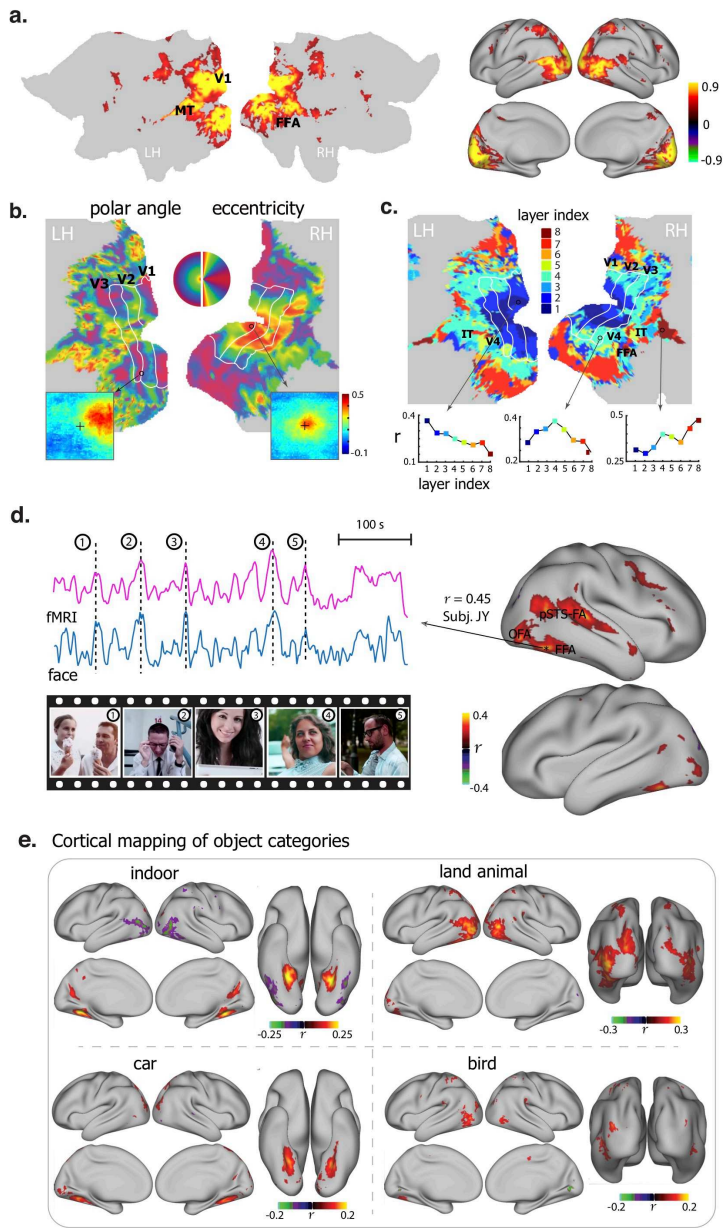

**Figure. 2. Functional alignment between the visual cortex and the CNN during natural vision (a) Cortical activation**. The maps show the cross correlations between the fMRI signals obtained during two repetitions of the identical movie stimuli. **(b) Retinotopic mapping**. Cortical representations of the polar angle (left) and eccentricity (right), quantified for the receptive-field center of every cortical location, are shown on the flattened cortical surfaces. The bottom insets show the receptive fields of two example locations from V1 (right) and V3 (left). The V1/V2/V3 borders defined from conventional retinotopic mapping are overlaid for comparison. **(c) Hierarchical mapping**. The map shows the index to the CNN layer most correlated with every cortical location. For three example locations, their correlations with different CNN layers are displayed in the bottom plots. **(d) Co-activation of FFA in the brain and the "Face" unit in the CNN**. The maps on the right show the correlations between cortical activity and the output time series of the "Face" unit in the 8th layer of CNN. On the left, the fMRI signal at a single voxel within the FFA is shown in comparison with the activation time series of the "Face" unit. Movie frames are displayed at five peaks co-occurring in both time series for one segment of the training movie. The selected voxel was chosen since it had the highest correlation with the "face" unit for other segments of the training movie, different from the one shown in this panel. **(e) Cortical mapping of other four categories**. The maps show the correlation between the cortical activity and the outputs of the 8th-layer units labeled as "indoor objects", "land animals", "car", "bird". See Supplementary Fig. 2, 3, 4 for related results from individual subjects.

For inter-subject encoding, we used the encoding models trained with data from one subject to predict another subject's cortical fMRI responses to the testing movie. The accuracy of inter-subject encoding was evaluated in the same way as done for intra-subject encoding (i.e. training and testing encoding models with data from the same subject). For inter-subject decoding, we used the decoding models trained with one subject's data to decode another subject's fMRI activity for reconstructing and categorizing the testing movie. The performance of inter-subject decoding was evaluated in the same way as for intra-subject decoding (i.e. training and testing decoding models with data from the same subject).

## Results

### Functional alignment between CNN and visual cortex

For exploring and modeling the relationships between the CNN and the brain, we used 374 video clips to constitute a training movie, presented twice to each subject for fMRI acquisition. From the training movie, the CNN extracted visual features through hundreds of thousands of units, which were organized into eight layers to form a trainable bottom-up network architecture (Supplementary Fig. 1). That is, the output of one layer was the input to its next layer. After the CNN was trained for image categorization (Krizhevsky et al., 2012), each unit encoded a particular feature through its weighted connections to its lower layer, and its output reported the representation of the encoded feature in the input image. The 1$^{st}$ layer extracted local features (e.g. orientation, color, contrast) from the input image; the 2$^{nd}$ through 7$^{th}$ layers extracted features with increasing nonlinearity, complexity, and abstraction; the highest layer reported the categorization probabilities (Krizhevsky et al., 2012; LeCun et al., 2015; Yamins and DiCarlo, 2016). See **Convolutional Neural Network** in **Methods** for details.

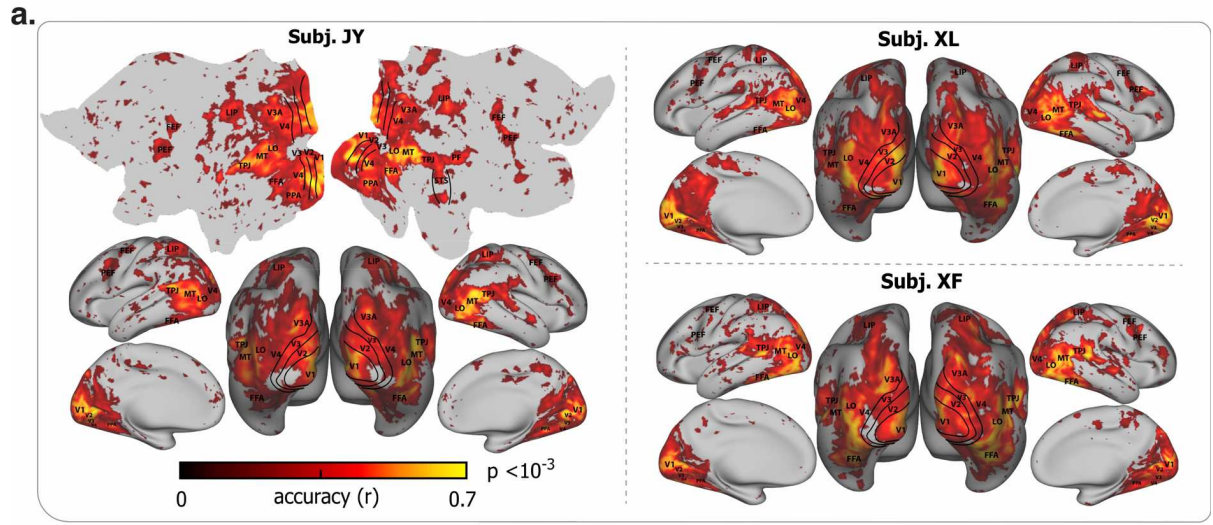

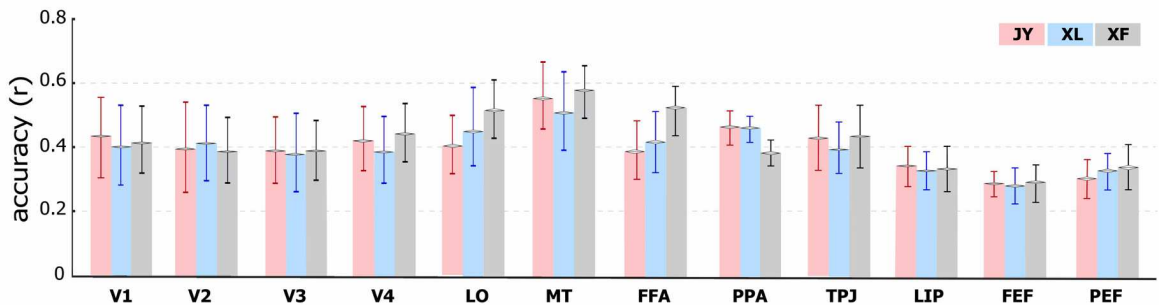

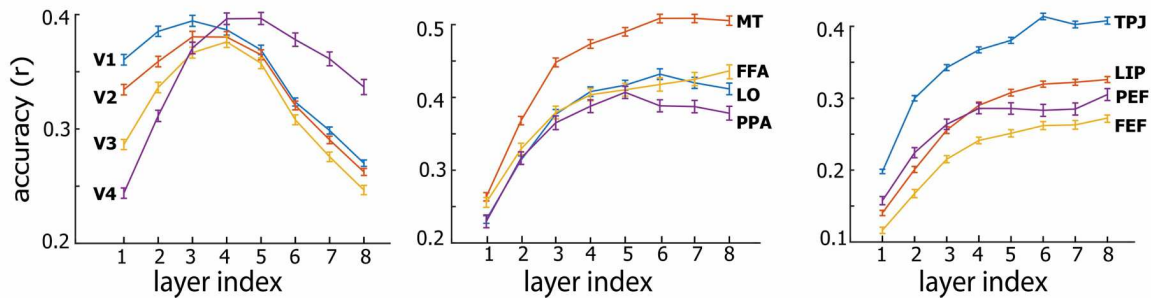

**Figure. 3. Cortical predictability given voxel-wise encoding models. (a)** Accuracy of voxel-wise encoding models in predicting the cortical responses to novel natural movie stimuli, which is quantified as the Pearson correlation between the measured and the model-predicted responses during the testing movie. **(b)** Prediction accuracy within regions of interest (ROIs) for three subjects. For each ROI, the prediction accuracy is summarized as the mean±std correlation for all voxels within the ROI. **(c)** Prediction accuracy for different ROIs by different CNN layers. For each ROI, the prediction accuracy was averaged across voxels within the ROI, and across subjects. The curves represent the mean, and the error bars stand for the standard error.

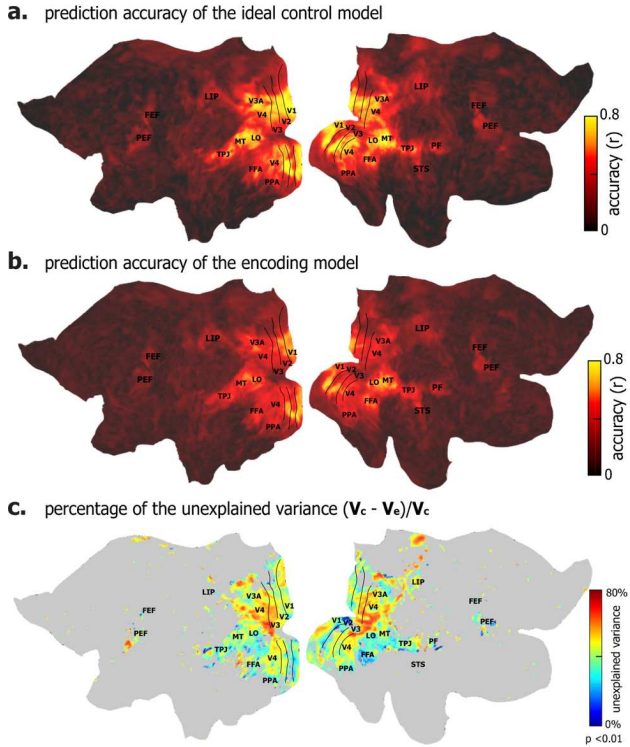

**Figure. 4. Explained variance of the encoding models. (a)** Prediction accuracy of the ideal control model (average across subjects). It defines the potentially explainable variance in the fMRI signal. **(b)** Prediction accuracy of the CNN-based encoding models (average across subjects). **(c)** The percentage of the explainable variance that is not explained by the encoding model. $V_c$ denotes the potentially explainable variance and $V_e$ denotes the variance explained by the encoding model. Note that this result was based on movie-evoked responses averaged over five repetitions of the testing movie, while the other five repetitions were used to define the ideally explainable variance. This was thus distinct from other figures, which were based on the responses averaged over all ten repetitions of the testing movie.

The hierarchical architecture and computation in the CNN appeared similar to the feedforward processing in the visual cortex (Yamins and DiCarlo, 2016). This motivated us to ask whether individual cortical locations were functionally similar to different units in the CNN given the training movie as the common input to both the brain and the CNN. To address this question, we first mapped the cortical activation with natural vision by evaluating the intra-subject reproducibility of fMRI activity when the subjects watched the training movie for the first vs. second time (Hasson et al., 2004; Lu et al., 2016). The resulting cortical activation was widespread over the entire visual cortex (Fig. 2.a) for all subjects (Supplementary Fig. 2). Then, we examined the relationship between the fMRI signal at every activated location and the output time series of every unit in the CNN. The latter indicated the time-varying representation of a particular feature in every frame of the training movie. The feature time series from each unit was log-transformed and convolved with the HRF, and then its correlation to each voxel's fMRI time series was calculated.

This bivariate correlation analysis was initially restricted to the $1^{st}$ layer in the CNN. Since the $1^{st}$-layer units filtered the image patches with a fixed size at a variable location, their correlations with a voxel's fMRI signal revealed its population receptive field (pRF) (see **Retinotopic mapping** in **Methods**). The bottom insets in Fig. 2.b. show the putative pRF of two example locations corresponding to peripheral and central visual fields. The retinotopic property was characterized by the polar angle and eccentricity of the center of every voxel's pRF (Supplementary Fig. 3.a), and mapped on the cortical surface (Fig. 2.b). The resulting retinotopic representations were consistent across subjects (Supplementary Fig. 3), and similar to the maps obtained with standard retinotopic mapping (Wandell et al., 2007; Abdollahi et al., 2014). The retinotopic organization reported here appeared more reasonable than the results obtained with a similar analysis approach but with natural picture stimuli (Eickenberg et al., 2016), suggesting an advantage of using movie stimuli for retinotopic mapping than using static pictures. Beyond retinotopy, we did not observe any orientation-selective representations (i.e. orientation columns), most likely due to the low spatial resolution of the fMRI data.

Extending the above bivariate analysis beyond the $1^{st}$-layer of the CNN, different cortical regions were found to be preferentially correlated with distinct layers in the CNN (Fig. 2.c). The lower to higher level features encoded by the $1^{st}$ through $8^{th}$ layers in the CNN were gradually mapped onto areas from the striate to extrastriate cortex along both ventral and dorsal streams (Fig. 2.c), consistently across subjects (Supplementary Fig. 4). These results agreed with findings from previous studies obtained with different analysis methods and static picture stimuli (Güçlü and van Gerven, 2015a,b; Khaligh-Razavi et al., 2016; Cichy et al., 2016; Eickenberg et al., 2016). We extended these findings to further show that the CNN could map the hierarchical stages of feedforward processing underlying dynamic natural vision, with a rather simple and effective analysis method.

Furthermore, an investigation of the categorical features encoded in the CNN revealed a close relationship with the known properties of some high-order visual areas. For example, a unit labeled as "face" in the output layer of the CNN was significantly correlated with multiple cortical areas (Fig. 2.d, right), including the fusiform face area (FFA), the occipital face area (OFA), and the face-selective area in the posterior superior temporal sulcus (pSTS-FA), all of which have been shown to contribute to face processing (Bernstein and Yovel, 2015). Such correlations were also relatively stronger on the right hemisphere than on the left hemisphere, in line with the right hemispheric dominance observed in many face-specific functional localizer experiments (Rossion et al., 2012). In addition, the fMRI response at the FFA and the output of the 'face' unit both showed notable peaks coinciding with movie frames that included human faces (Fig. 2.d, left). These results exemplify the utility of mapping distributed neural-network representations of object categories automatically detected by the CNN. In this sense, it is more convenient than doing so by manually labeling movie frames, as in prior studies (Russ and Leopold, 2015; Huth et al., 2012). Similar strategies were also used to reveal the network representations of 'indoor scenes', 'land animals', 'car', and 'bird' (Fig. 2.e).

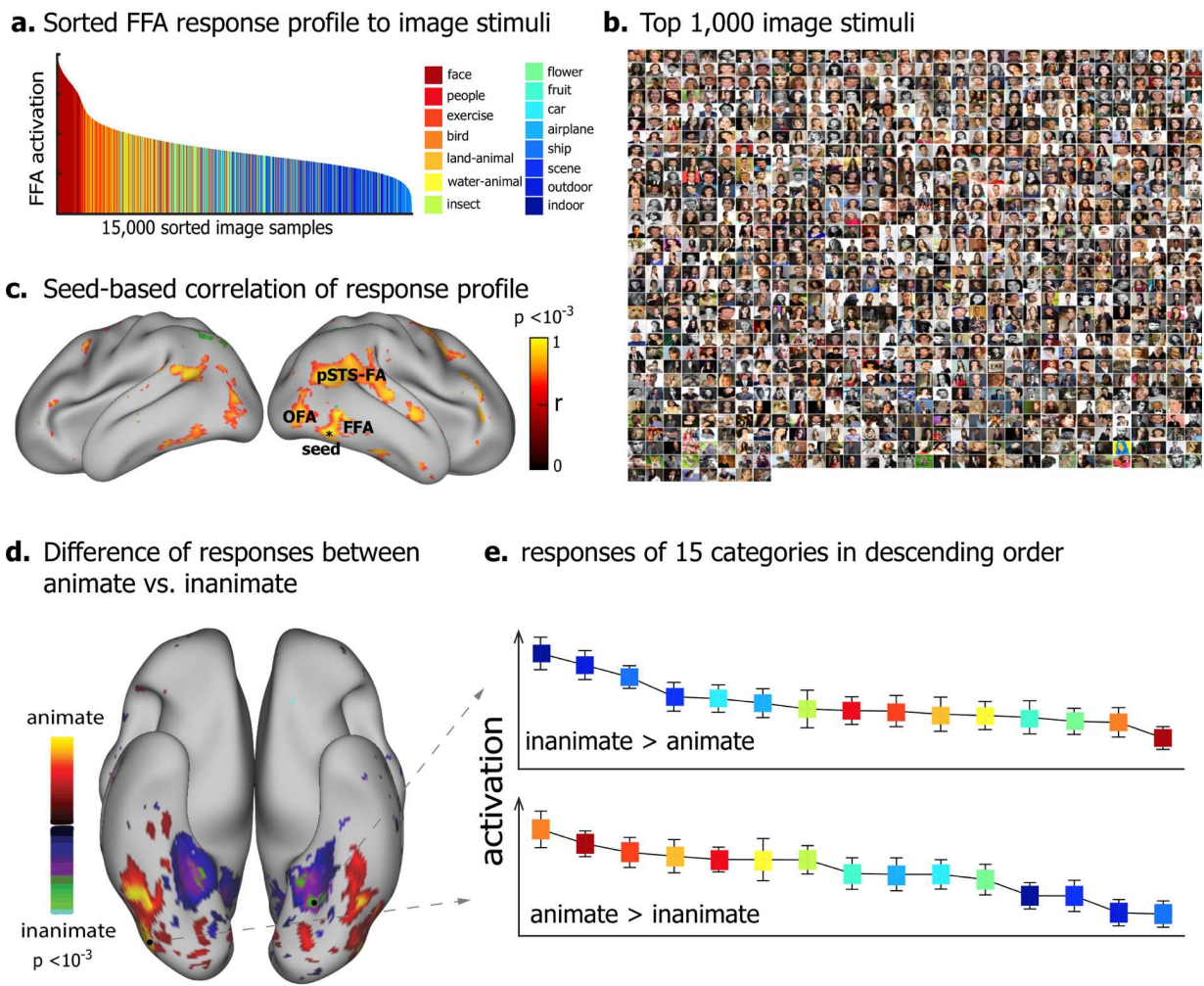

**Figure. 5. Cortical representations of single-pictures or categories. (a)** The model-predicted response profile at a selected voxel in FFA given 15,000 natural pictures from 15 categories, where the selected voxel had the highest prediction accuracy when the encoding model was evaluated using the testing movie. The voxel's responses are sorted in descending order. **(b)** The top 1,000 pictures that generate the greatest responses at this FFA voxel. **(c)** Correlation of the response profile at this "seed" voxel with those at other voxels (p<0.001, Bonferroni correction). **(d)** The contrast between animate versus inanimate pictures in the model-predicted responses (two-sample t-test, p<0.001, Bonferroni correction). **(e)** The categorical responses at two example voxels. These two voxels show the highest animate and inanimate responses, respectively. The colors correspond to the categories in (a). The results are from Subject JY, see Supplementary Fig. 5 for related results from other subjects.

Taken together, the above results suggest that the hierarchical layers in the CNN implement similar computational principles as cascaded visual areas along the brain's visual pathways. The CNN and the visual cortex not only share similar representations of some low-level visual features (e.g. retinotopy) and high-level semantic features (e.g. face), but also share similarly hierarchical representations of multiple intermediate levels of progressively abstract visual information (Fig. 2).

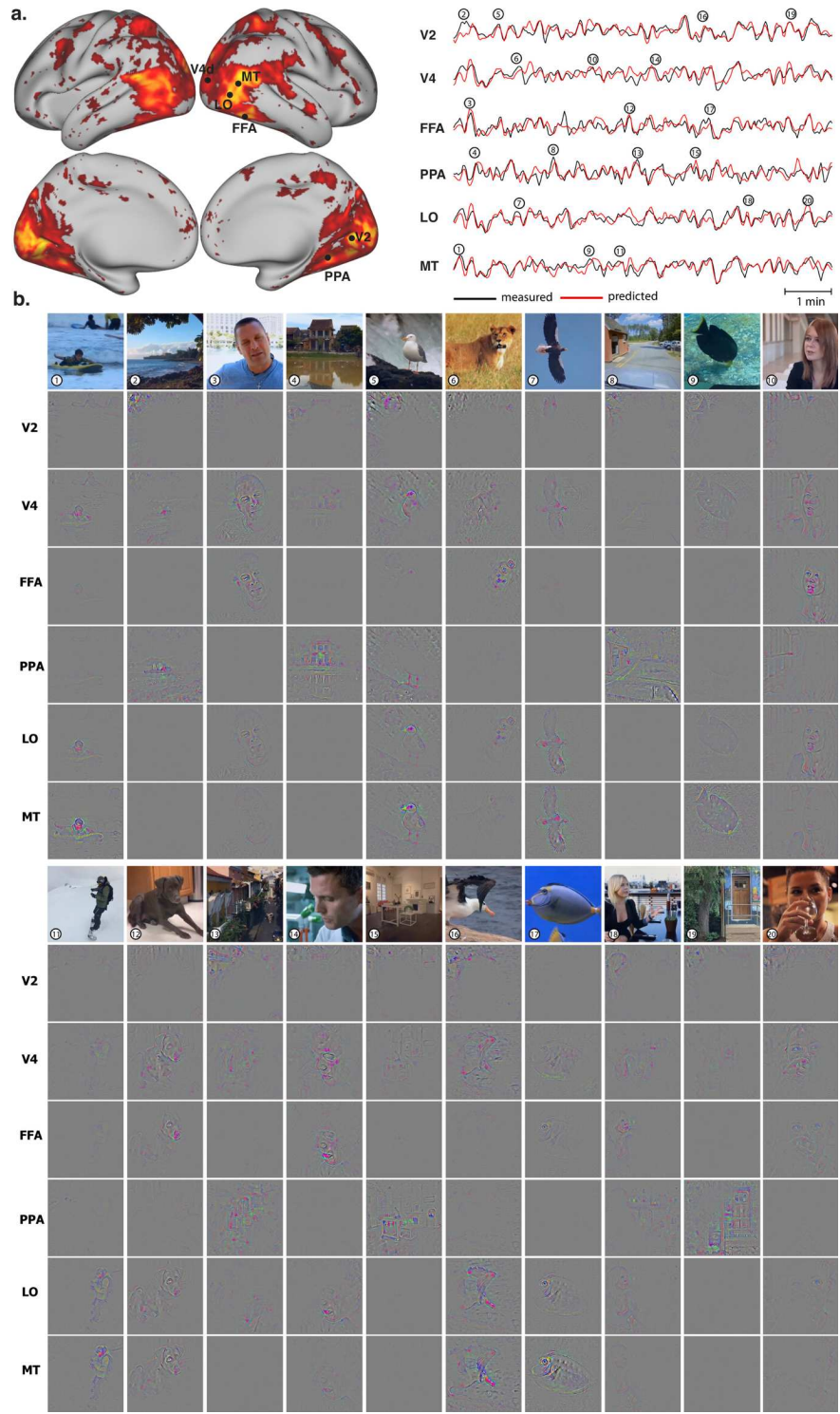

**Figure. 6. Neural encoding models predict cortical responses and visualize functional representations at individual cortical locations. (a)** Cortical predictability for subject JY, same as Fig. 3a. The measured (black) and predicted (red) response time series are also shown in comparison for six locations at V2, V4, LO, MT, PPA and FFA. For each area, the selected location was the voxel within the area where the encoding models yielded the highest prediction accuracy during the testing movie **(b)** Visualizations of the 20 peak responses at each of the six locations shown in (a). The presented movie frames are shown in the top row, and the corresponding visualizations at six locations are shown in the following rows. The results are from Subject JY, see Supplementary Fig. 6 and 7 for related results from other subjects.

**Neural encoding**

Given the functional alignment between the human visual cortex and the CNN as demonstrated above and previously by others (Güçlü and van Gerven, 2015a; Cichy et al., 2016; Eickenberg et al., 2016), we further asked whether the CNN could be used as a predictive model of the response at any cortical location given any natural visual input. In other words, we attempted to establish a voxel-wise encoding model (Kay et al., 2008; Naselaris et al., 2011) by which the fMRI response at each voxel was predicted from the output of the CNN. Specifically, for any given voxel, we optimized a linear regression model to combine the outputs of the units from a single layer in CNN to best predict the fMRI response during the training movie. We identified and used the principal components of the CNN outputs as the regressors to explain the fMRI voxel signal. Given the training movie, the output from each CNN layer could be largely explained by much fewer components. For the $1^{st}$ through $8^{th}$ layers, 99% of the variance in the outputs from 290400, 186624, 64896, 64896, 43264, 4096, 4096, 1000 units could be explained by 10189, 10074, 9901, 10155, 10695, 3103, 2804, 241 components, respectively. Despite dramatic dimension reduction especially for the lower layers, information loss was negligible (1%), and the reduced feature dimension largely mitigated overfitting when training the voxel-wise encoding model.

After training a separate encoding model for every voxel, we used the models to predict the fMRI responses to five 8-min testing movies. These testing movies included different video clips from those in the training movie, and thus unseen by the encoding models to ensure unbiased model evaluation. The prediction accuracy ($r$), measured as the correlation between the predicted and measured fMRI responses, was evaluated for every voxel. As shown in Fig. 3.a, the encoding models could predict cortical responses with reasonably high accuracies for nearly the entire visual cortex, much beyond the spatial extent predictable with low-level visual features (Nishimoto et al. 2011) or high-level semantic features (Huth et al., 2012) alone. The model-predictable cortical areas shown in this study also covered a broader extent than was shown in prior studies using similar CNN-based feature models (Güçlü and van Gerven, 2015a; Eickenberg et al., 2016). The predictable areas even extended beyond the ventral visual stream, onto the dorsal visual stream, as well as areas in parietal, temporal, and frontal cortices (Fig. 3.a). These results suggest that object representations also exist in the dorsal visual stream, in line with prior studies (Freud et al., 2016; de Haan and Cowey, 2011).

Regions of interest (ROI) were selected as example areas in various levels of visual hierarchy: V1, V2, V3, V4, lateral occipital (LO), middle temporal (MT), fusiform face area (FFA), parahippocampal place area (PPA), lateral intraparietal (LIP), temporo-parietal junction (TPJ), premotor eye field (PEF), and frontal eye field (FEF). The prediction accuracy, averaged within each ROI, was similar across subjects, and ranged from 0.4 to 0.6 across the ROIs within the visual cortex and from 0.25 to 0.3 outside the visual cortex (Fig. 3.b). These results suggest that the internal representations of the CNN explain cortical representations of low, middle, and high-level visual features to similar degrees. Different layers in the CNN contributed differentially to the prediction at each ROI (Fig. 3.c). Also see Fig. 6.a for the comparison between the predicted and measured fMRI time series during the testing movie at individual voxels.

Although the CNN-based encoding models predicted partially but significantly the widespread fMRI responses during natural movie viewing, we further asked where and to what extent the models failed to fully predict the movie-evoked responses. Also note that the fMRI measurements contained noise and reflected in part spontaneous activity unrelated to the movie stimuli. In the presence of the noise, we defined a control model, in which the fMRI signal averaged over five repetitions of the testing movie was used to predict the fMRI signal averaged over the other five repetitions of the same movie. This control model served to define the explainable variance for the encoding model, or the ideal prediction accuracy (Fig. 4.a), against which the prediction accuracy of the encoding models (Fig. 4.b) was compared. Relative to the explainable variance, the CNN model tended to be more predictive of ventral visual areas (Fig. 4.c), which presumably sub-served the similar goal of object recognition as did the CNN (Yamins and DiCarlo, 2016). In contrast, the CNN model still fell relatively short in predicting the responses along the dorsal pathway (Fig. 4.c), likely because the CNN did not explicitly extract temporal features that are important for visual action (Hasson et al., 2008).

**Cortical representations of single-pictures or categories**

The voxel-wise encoding models provided a fully computable pathway through which any arbitrary picture could be transformed to the stimulus-evoked fMRI response at any voxel in the visual cortex. As initially explored before (Eickenberg et al. 2016), we conducted a high-throughput "virtual-fMRI" experiment with 15,000 images randomly and evenly sampled from 15 categories in ImageNet (Deng et al., 2009; Russakovsky et al.,

2015). These images were taken individually as input to the encoding model to predict their corresponding cortical fMRI responses. As a result, each voxel was assigned with a predicted response to every picture, and its response profile across individual pictures reported the voxel's functional representation (Mur et al., 2012). For an initial proof of concept, we selected a single voxel that showed the highest prediction accuracy within FFA – an area for face recognition (Kanwisher et al., 1997; Bernstein and Yovel, 2015; Rossion et al., 2012). This voxel's response profile, sorted by the response level, showed strong face selectivity (Fig. 5.a). The top 1,000 pictures that generated the strongest responses at this voxel were mostly human faces (94.0%, 93.9%, and 91.9%) (Fig. 5.b). Such a response profile was not only limited to the selected voxel, but shared across a network including multiple areas from both hemispheres, e.g. FFA, OFA, and pSTS-FA (Fig. 5c). It demonstrates the utility of the CNN-based encoding models for analyzing the categorical representations in voxel, regional, and network levels. Extending from this example, we further compared the categorical representation of every voxel, and generated a contrast map for the differential representations of animate vs. inanimate categories (Fig. 5d). We found that the lateral and inferior temporal cortex (including FFA) was relatively more selective to animate categories, whereas the parahippocampal cortex was more selective to inanimate categories (Fig. 5.d), in line with previous findings (Kriegeskorte et al., 2008; Naselaris et al., 2012). Supplementary Fig. S5 shows the comparable results from the other two subjects.

**Visualizing single-voxel representations given natural visual input**

Not only could the voxel-wise encoding models predict how a voxel responded to different pictures or categories, such models were also expected to reveal how different voxels extract and process different visual information from the same visual input. To this end, we developed a method to visualize for each single voxel its representation given a known visual input. The method was to identify a pixel pattern from the visual input that accounted for the voxel response through the encoding model, revealing the voxel's representation of the input.

To visualize single-voxel representations, we selected six voxels from V2, V4, LO, MT, FFA and PPA (as shown in Fig. 6.a, left) as example cortical locations at different levels of visual hierarchy. For these voxels, the voxel-wise encoding models could well predict their individual responses to the testing movie (Fig. 6.a, right). At 20 time points when peak responses were observed at one or multiple of these voxels, the visualized representations shed light on their different functions (Fig. 6). It was readily notable that the visual representations of the V2 voxel were generally confined to a fixed part of the visual field, and showed pixel patterns with local details; the V4 voxel mostly extracted and processed information about foreground objects rather than from the background; the MT voxel selectively responded to the part of the movie frames that implied motion or action; the LO voxel represented either body parts or facial features;

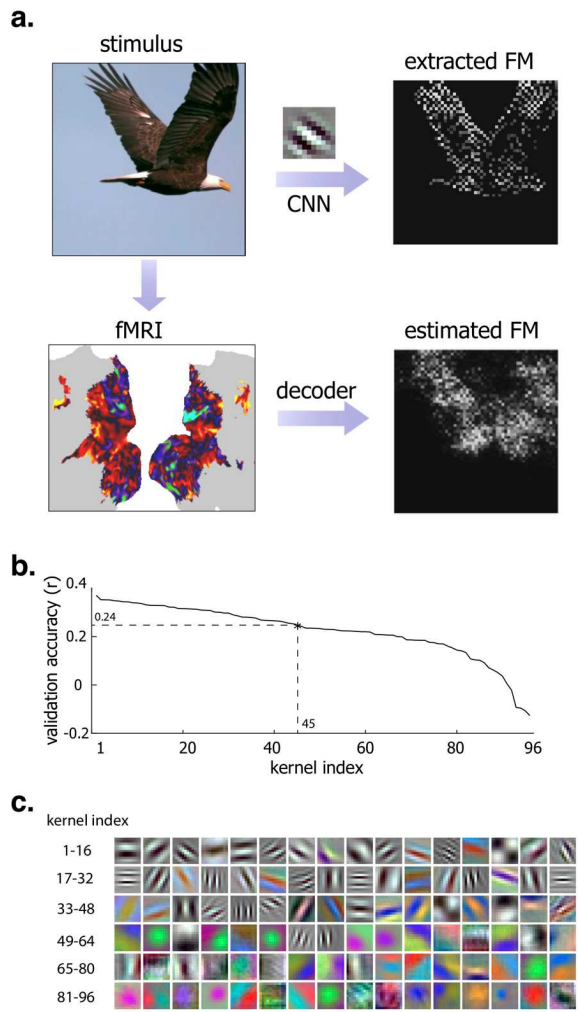

**Figure 7. fMRI-based estimation of the first-layer feature maps (FM).** (a) For each movie frame, the feature maps extracted from the kernels in the first CNN layer were estimated from cortical fMRI data through decoders trained with the training movie. For an example movie frame (flying eagle) in the testing movie, its feature map extracted with an orientation-coded kernel revealed the image edges. In comparison, the feature map estimated from fMRI was similar, but blurrier. (b) The estimation accuracy for all 96 kernels, given cross-validation within the training data. The accuracies were ranked and plotted from the highest to lowest. Those kernels with high accuracies (r > 0.24) were selected and used for reconstructing novel natural movies in the testing phase. (c) 96 kernels in the first layer are ordered in a descending manner according to their cross-validation accuracy.

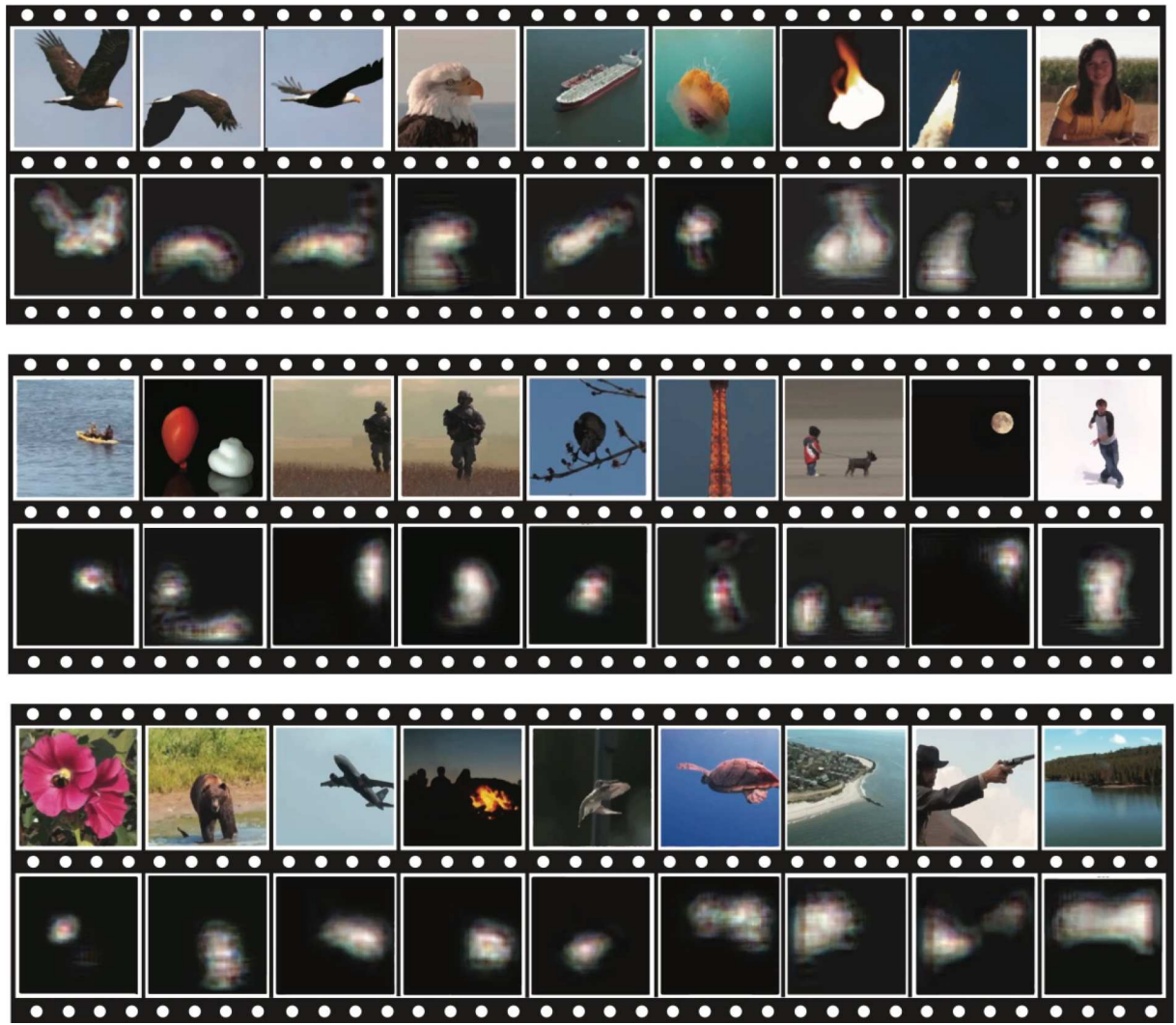

**Figure. 8. Reconstruction of a dynamic visual experience**. For each row, the top shows the example movie frames seen by one subject; the bottom shows the reconstruction of those frames based on the subject's cortical fMRI responses to the movie. See Movie 1 for the reconstructed movie.

the FFA voxel responded selectively to human and animal faces, whereas the PPA voxel revealed representations of background, scenes, or houses. These visualizations offered intuitive illustration of different visual functions at different cortical locations, extending beyond their putative receptive-field size and location.

**Neural decoding**

While the CNN-based encoding models described the visual representations of individual voxels, it is the distributed patterns of cortical activity that gave rise to realistic visual and semantic experiences. To account for distributed neural coding, we sought to build a set of decoding models that combine individual voxel responses in a way to reconstruct the visual input to the eyes (visual reconstruction), and to deduce the visual percept in the mind (semantic categorization). Unlike previous studies (Haxby et al., 2001; Carlson et al., 2002; Thirion et al., 2006; Kay et al., 2008; Nishimoto et al., 2011), our strategy for decoding was to establish a computational path to directly transform fMRI activity patterns onto individual movie frames and their semantics captured at the fMRI sampling times.

For visual reconstruction, we defined and trained a set of multivariate linear regression models to combine the fMRI signals across cortical voxels (not confined to V1, but all in Supplementary S2.e) in an optimal way to match every feature map in the $1^{st}$ CNN layer during the training movie. Such feature maps resulted from extracting various local features from every frame of the training movie (Fig. 7.a). By 20-fold cross-validation within the training data, the models tended to give more reliable estimates for 45 (out of 96) feature maps (Fig. 7.b), mostly related to features for detecting orientations and edges, whereas the estimates were less reliable for most color features (Fig. 7.c). In the testing phase, the trained models were used to convert distributed cortical responses generated by the testing movie to the estimated feature maps for the $1^{st}$-layer features. The reconstructed feature maps were found to be correlated with the actual feature maps directly extracted by the CNN (r=0.30±0.04). By using the De-CNN, every estimated feature map was transformed back to the pixel space, where they were combined to reconstruct the individual frames of the testing movie. Fig. 8 shows some examples of the movie frames reconstructed from fMRI vs. those actually presented. The reconstruction clearly captured the location, shape, and motion of salient objects, despite missing color. Perceptually less salient objects and the background were poorly reproduced in the reconstructed images. Such predominance of foreground objects is likely attributed to the effects of visual salience and attention on fMRI activity (Itti and Koch, 2001; Desimore and Duncan, 1995). Thus, the decoding in this study does not simply invert retinotopy (Thirion et al., 2006) to reconstruct the original image, but tends to reconstruct the image parts relevant to visual perception. Miyawaki et al. previously used a similar computational strategy for direct reconstruction of simple pixel patterns, e.g. letters and shapes, with binary-valued local image bases (Miyawaki et al., 2008). In contrast to the method in that study, the decoding method in this study utilized data-driven and biologically-relevant visual features to better account for natural image statistics (Olshausen and Field, 1996; Hyvarien et al., 2009). In addition, the decoding models, when trained and tested with natural movie stimuli, represented an apparently better account of cortical activity underlying natural vision, than the model trained with random images and tested for small-sized artificial stimuli (Miyawaki et al., 2008).

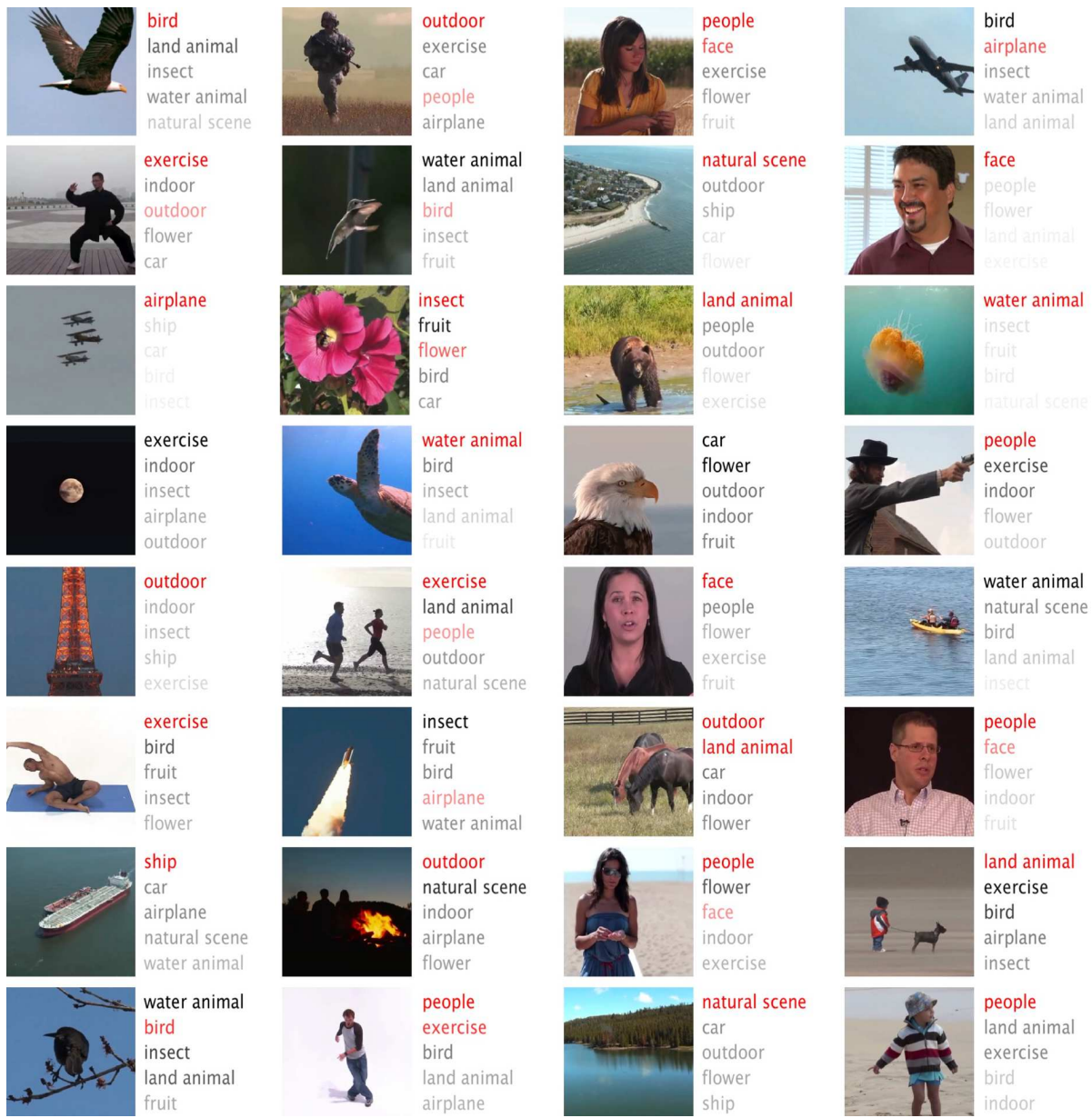

**Figure. 9. Semantic categorization of natural movie stimuli**. For each movie frame, the top five categories determined from cortical fMRI activity are shown in the order of descending probabilities from the top to the bottom. The probability is also color coded in the gray scale with the darker gray indicative of higher probability. For comparison, the true category labeled by the subjects is shown in red. Here, we present the middle frame of every continuous video clip in the testing movie that could be labeled as one of the pre-defined categories. See Movie 1 for all other frames.

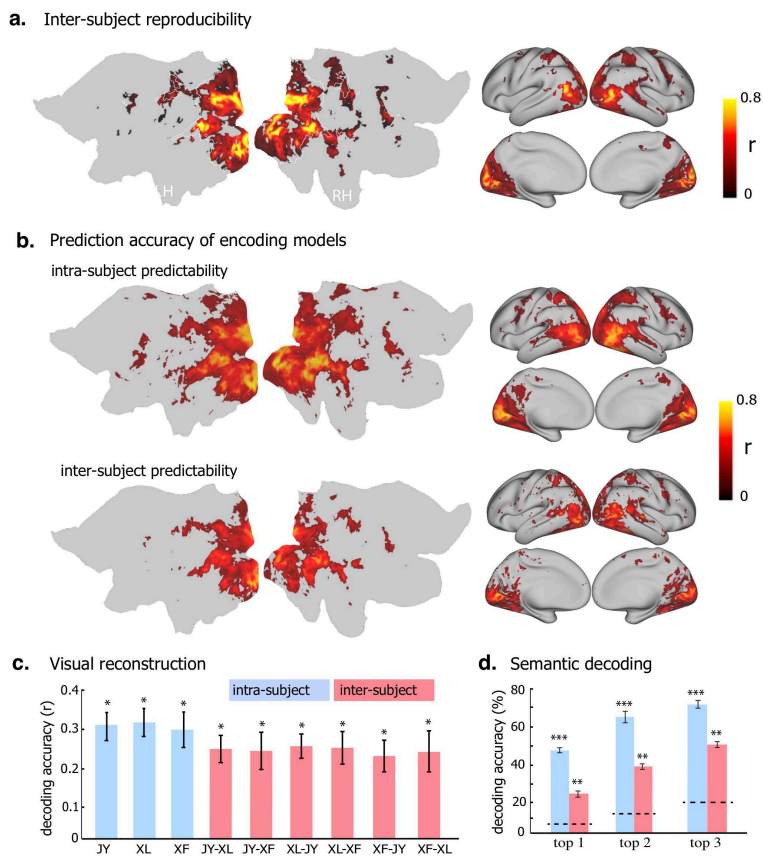

**Figure. 10. Encoding and decoding within vs. across subjects.** (a) Average inter-subject reproducibility of fMRI activity during natural stimuli. (b) Cortical response predictability with the encoding models trained and tested for the same subject (i.e. intra-subject encoding) or for different subjects (i.e. inter-subject encoding). (c) Accuracy of visual reconstruction by intra-subject (blue) vs. inter-subject (red) decoding for one testing movie. The y axis indicates the spatial cross correlation between the fMRI-estimated and CNN-extracted feature maps for the $1^{st}$ layer in the CNN. The x axis shows multiple pairs of subjects (JY, XL and XF). The first subject indicates the subject from whom the decoder was trained; the second subject indicates the subject for whom the decoder was tested. (d) Accuracy of categorization by intra-subject (blue) vs. inter-subject (red) decoding. The top-1, top-2 and top-3 accuracy indicates the percentage by which the true category is within the $1^{st}$, $2^{nd}$, and $3^{rd}$ most probable categories predicted from fMRI, respectively. For both (c) and (d), the bar height indicates the average prediction accuracy; the error bar indicates the standard error of the mean; the dashed lines are chance levels. (*$p<10^{-4}$, **$p<10^{-10}$, ***$p<10^{-50}$). See Movie 2 for the reconstructed movie on the basis of inter-subject decoding.

To identify object categories from fMRI activity, we optimized a decoding model to estimate the category that each movie frame belonged to. Briefly, the decoding model included two parts: 1) a multivariate linear regression model that used the fMRI signals to estimate the semantic representation in the $7^{th}$ (i.e. the $2^{nd}$-highest) CNN layer, 2) the built-in transformation from the $7^{th}$ to the $8^{th}$ (or output) layer in the CNN, to estimate the categorization probabilities from the decoded semantic representation. The first part of the model was trained with the fMRI data during the training movie; the second part was established by retraining the CNN for image classification into 15 categories. After training, we evaluated the decoding performance with the testing movie. Fig. 9 shows the top-5 decoded categories, ordered by their descending probabilities, in comparison with the true categories shown in red. On average, the top-1/top-2/top-3 accuracies were about 48%/65%/72%, significantly better than the chance levels (6.9%/14.4%/22.3%) (Table 1). These results confirm that cortical fMRI activity contained rich categorical representations, as previously shown elsewhere (Huth et al., 2012, 2016a, 2016b). Along with visual reconstruction, direct categorization yielded textual descriptions. As an example, a flying bird seen by a subject was not only reconstructed as a bird-

**Decoding accuracy for the semantic descriptions of a novel movie**

|  | train \ test | subject 1 | subject 2 | subject 3 |
|---|---|---|---|---|
| top 1 | subject 1 | 42.52% (91/214) | 24.30% (52/214) | 23.83% (51/214) |
|  | subject 2 | 20.09% (43/214) | 50.47% (108/214) | 22.90% (49/214) |
|  | subject 3 | 24.77% (53/214) | 33.64% (72/214) | 50.00% (107/214) |
| top 2 | subject 1 | 59.81% (128/214) | 41.12% (88/214) | 43.93% (94/214) |
|  | subject 2 | 35.51% (76/214) | 70.09% (150/214) | 35.98% (77/214) |
|  | subject 3 | 41.12% (88/214) | 42.06% (90/214) | 66.36% (142/214) |
| top 3 | subject 1 | 67.76% (145/214) | 55.14% (118/214) | 53.27% (114/214) |
|  | subject 2 | 48.13% (103/214) | 74.77% (160/214) | 50.93% (109/214) |
|  | subject 3 | 50.93% (109/214) | 52.34% (112/214) | 72.90% (156/214) |

**Table 1.** Three sub-tables show the top-1, top-2 and top-3 accuracies of categorizing individual movie frames by using decoders trained with data from the same (intra-subject) or different (inter-subject) subject. Each row shows the categorization accuracy with the decoder trained with a specific subject's training data; each column shows the categorization accuracy with a specific subject's testing data and different subjects' decoders. The accuracy was quantified as the percentage by which individual movie frames were successfully categorized as one of the top-1, top-2, or top-3 categories. The accuracy was also quantified as a fraction number (shown next to the percentage number): the number of correctly categorized frames over the total number of frames that could be labeled by the 15 categories (N=214 for one 8-min testing movie).

like image, but also described as a word "bird" (see the first frame in Figs. 8 & 9).

**Cross-subject encoding and decoding**

Different subjects' cortical activity during the same training movie were generally similar, showing significant inter-subject reproducibility of the fMRI signal ($p<0.01$, t-test, Bonferroni correction) for 82% of the locations within visual cortex (Fig. 10.a). This lent support to the feasibility of neural encoding and decoding across different subjects – predicting and decoding one subject's fMRI activity with the encoding/decoding models trained with data from another subject. Indeed, it was found that the encoding models could predict cortical fMRI responses across subjects with still significant, yet reduced, prediction accuracies for most of the visual cortex (Fig. 10.b). For decoding, low-level feature representations (through the 1$^{st}$ layer in the CNN) could be estimated by inter-subject decoding, yielding reasonable accuracies only slightly lower than those obtained by training and testing the decoding models with data from the same subject (Fig. 10.c). The semantic categorization by inter-subject decoding yielded top-1 through top-3 accuracies as 24.9%, 40.0% and 51.8%, significantly higher than the chance levels (6.9%, 14.4% and 22.3%), although lower than those for intra-subject decoding (47.7%, 65.4%, 71.8%) (Fig. 10.d and Table 1). Together, these results provide evidence for the feasibility of establishing neural encoding and decoding models for a general population, while setting up the baseline for potentially examining the disrupted coding mechanism in pathological conditions.

## Discussion

This study extends a growing body of literature in using deep learning models for understanding and modeling cortical representations of natural vision (Khaligh-Razavi and Kriegeskorte, 2014; Yamins et al., 2014; Güçlü and van Gerven, 2015a,b; Cichy et al., 2016; Kubilius et al., 2016; Horikawa & Kamitani, 2017; Eickenberg et al., 2016). In particular, it generalizes the use of convolutional neural network to explain and decode widespread fMRI responses to naturalistic movie stimuli, extending the previous findings obtained with static picture stimuli. This finding lends support to the notion that cortical activity underlying dynamic natural vision is largely shaped by hierarchical feedforward processing driven towards object recognition, not only for the ventral stream, but also for the dorsal stream, albeit to a lesser degree. It sheds light on the object representations along the dorsal stream.

Despite its lack of recurrent or feedback connections, the CNN enables a fully computable predictive model of cortical representations of any natural visual input. The voxel-wise encoding model enables the visualization of single-voxel representation, to reveal the distinct functions of individual cortical locations during natural vision. It further creates a high-throughput computational workbench for synthesizing cortical responses to natural pictures, to enable cortical mapping of category representation and selectivity without running fMRI experiments. In addition, the CNN also enables direct decoding of cortical fMRI activity to estimate the feature representations in both visual and semantic spaces, for real-time visual reconstruction and semantic categorization of natural movie stimuli. In summary, the CNN-based encoding and decoding models, trained with hours of fMRI data during movie viewing, establish a computational account of feedforward cortical activity throughout the entire visual cortex and across all levels of processing. Subsequently, we elaborate the implications from methodology, neuroscience, and artificial intelligence perspectives.

**CNN predicts nonlinear cortical responses throughout the visual hierarchy**

The brain segregates and integrates visual input through cascaded stages of processing. The relationship between the visual input and the neural response bears a variety of nonlinearity and complexity (Yamins and DiCarlo, 2016). It is thus impossible to hand-craft a general class of models to describe the neural code for every location, especially for those involved in the mid-level processing. The CNN accounts for natural image statistics with a hierarchy of nonlinear feature models learned from millions of labeled images. The feature representations of any image or video can be automatically extracted by the CNN, progressively ranging from the visual to semantic space. Such feature models offer a more convenient and comprehensive set of predictors to explain the evoked fMRI responses, than are manually defined (Russ and Leopold, 2015; Huth et al., 2012). For each voxel, the encoding model selects a subset from the feature bank to best match the voxel response with a linear projection. This affords the flexibility to optimally model the nonlinear stimulus-response relationship to maximize the response predictability for each voxel.

In this study, the model-predictable voxels cover nearly the entire visual cortex (Fig. 3.a), much beyond the early visual areas predictable with Gabor or motion filters (Daugman, 1985; Kay et al., 2008; Nishimoto et al., 2011), or with manually-defined categorical features (Huth et al., 2012; Russ and Leopold, 2015). It is also broader than the incomplete ventral stream previously predicted by similar models trained with limited static pictures (Güçlü and van Gerven, 2015a; Horikawa and Kamitani, 2017; Eickenberg et al., 2016). The difference is likely attributed to the larger sample size of our training data, conveniently afforded by video stimuli rather than picture stimuli. The PCA-based feature-dimension reduction also contributes to more robust and efficient model training. However, the encoding models only account for a fraction of the explainable variance (Fig. 4), and hardly explain the most lateral portion of early visual areas (Fig. 3.a). This area tends to have a lower SNR, showing lower intra-subject reproducibility (Fig. 2.a) or explainable variance (Fig. 4.a). The same issue also appears in other studies (Hasson et al., 2004; Güçlü and van Gerven, 2015a), whereas the precise reason remains unclear.

Both the ventral stream and the CNN are presumably driven by the same goal of object recognition (Yamins and DiCarlo, 2016). Hence, it is not surprising that the CNN is able to explain a significant amount of cortical activity along the ventral stream, in line with prior studies (Khaligh-Razavi and Kriegeskorte, 2014; Yamins et al., 2014; Güçlü and van Gerven, 2015a; Eickenberg et al., 2016). It further confirms the paramount role of feedforward processing in object recognition and categorization (Serre et al., 2007).

What is perhaps surprising is that the CNN also predicts dorsal-stream activity. The ventral-dorsal segregation is a classical principle of visual processing: the ventral stream is for perception ("what"), and the dorsal stream is for action ("where") (Goodale and Milner, 1992). As such, the CNN aligns with the former but not the latter. However, dorsal and ventral areas are interconnected, allowing cross-talk between the pathways (Schenk and McIntosh, 2010). The dichotomy of visual streams is debatable (de Haan and Cowey, 2011). Object representations exist in both ventral and dorsal streams with likely dissociable roles in visual perception (Freud et al., 2016). Our study supports this notion. The hierarchical features extracted by the CNN are also mapped onto the dorsal stream, showing a representational gradient of complexity, as does the ventral stream (Güçlü and van Gerven, 2015a). Nevertheless, the CNN accounts for a higher portion of the explainable variance for the ventral stream than for the dorsal stream (Fig. 4). We speculate that motion and attention sensitive areas in the dorsal stream require more than feedforward perceptual representations, while involving recurrent and feedback connections (Kafaligonul et al., 2015) that are absent in the CNN. In this regard, we would like to clarify that the CNN in the context of this paper is driven by image recognition and extracts spatial features, in contrast to 3-D convolutional network trained to extract spatiotemporal features for action recognition (Tran et al., 2014), which was another plausible model for the dorsal-stream activity (Güçlü and van Gerven, 2015b).

**Visualization of single-voxel representation reveals functional specialization**

An important contribution of this study is the method for visualizing single-voxel representation. It reveals the specific pixel pattern from the visual input that gives rise to the response at the voxel of interest. The method is similar to those for visualizing the representations of individual units in the CNN (Zeiler and Fergus, 2014; Springenberg et al., 2014). Extending from CNN units to brain voxels, it is helpful to view the encoding models as an extension of the CNN, where units are linearly projected onto voxels through voxel-wise encoding models. By this extension, the pixel pattern is optimized to maximize the model prediction of the voxel response, revealing the voxel's representation of the given visual input, using a combination of masking (Zhou et al., 2014; Li, 2016;) and gradient (Springenberg et al., 2014; Simonyan et al., 2013; Baehrens et al., 2010;) based methods. Here, visualization is tailored to each voxel, instead of each unit or layer in the CNN, setting it apart from prior studies (Zeiler and Fergus, 2014; Simonyan et al., 2013; Springenberg et al., 2014; Güçlü and van Gerven, 2015a).

Utilizing this visualization method, one may reveal the distinct representations of the same visual input at different cortical locations. As exemplified in Fig. 6, visualization uncovers the increasingly complex and category-selective representations for locations running downstream along the visual pathways. It offers intuitive insights into the distinct functions of different locations, e.g. the complementary representations at FFA and PPA. Although we focus on the methodology, our initial results merit future studies for more systematic characterization of the representational differences among voxels in various spatial scales. The visualization method is also applicable to single or multi-unit activity, to help understand the localized responses of neurons or neuronal ensembles (Yamins et al., 2014).

**High-throughput computational workbench for studying natural vision**

The CNN-based encoding models, trained with a large and diverse set of natural movie stimuli, can be generalized to other novel visual stimuli. Given this generalizability, one may use the trained encoding models to predict and analyze cortical responses to a large number of natural pictures or videos, much beyond what is practically doable with fMRI scans. As such, the encoding models constitute a high-throughput computational workbench for studying the neural representations of natural vision. As shown here and elsewhere (Eickenberg et al., 2016), this workbench is immediately usable for mapping categorical representation, contrast, and selectivity, to yield novel hypotheses for further experimental investigations. Open-access software platform is much desirable to further leverage this potential.

**Direct visual reconstruction of a natural movie**

For decoding cortical activity, the CNN enables direct reconstruction of natural movies. It does not require any comparison between the observed activity pattern and those generated by or predicted from candidate pictures. This sets our method apart from multivariate pattern analysis (Kamitani and Tong, 2005; Haynes and Rees, 2006; Norman et al., 2006) and encoding-model-based decoding (Kay et al., 2008; Naselaris et al., 2009; Nishimoto et al., 2011). In particular, Nishimoto et al. (2011) published the first, and to date the only, attempt to reconstruct natural movies. They used a "try-and-error" strategy: searching a huge prior set of videos for the most likely stimuli that would match the measured cortical activity through model prediction by the encoding models. Arguably, this strategy is difficult to scale up because it is impossible for any prior set to be fully inclusive. The identification or reconstruction accuracy is dependent on and biased by the samples in the prior set. The need for a large prior set is also computationally expensive, limiting the decoding efficiency.

A prior study (Miyawaki et al., 2008) tried to avoid these limitations. In that study, the fMRI signals were used to estimate the contrast of local image bases, which in turn were combined to directly reconstruct small, simple, and binary images. While the method is not constrained or biased by any image prior, binary image bases are not suitable for describing natural image statistics even in the lowest level (Olshausen and Field, 1996; Hyvarien et al., 2009). Also note that the decoding models in that study were trained with a small set of random images, and tested with simple letters and shapes. However, realistic visual input is complex and dynamic, and natural vision involves salience and attention (Itti and Koch, 2001; Desimore and Duncan, 1995). Such complexity is unlikely captured by random and binary pixel patterns (Lu et al., 2016). The overall strategy, as described in (Miyawaki et al., 2008), is not readily usable to decode dynamic natural visual experiences.

Our decoding method does not require any prior set of candidate images, setting itself apart from the encoding-model-based decoding (Nishimoto et al., 2011). It also uses features learned from natural images, different from the method in (Miyawaki et al., 2008). The latter is important because the features in the CNN are biologically relevant (Yamins and DiCarlo, 2016) and capture information useful for perception (LeCun et al., 2015). In particular, the $1^{st}$ layer includes features of orientation, contrast, edge, and color, forming a more informative basis set than binary image bases (Miyawaki et al., 2008).

In this study, visual reconstruction was only based on the fMRI-decoded $1^{st}$-layer features. Although the feature representations from other layers could also be estimated with comparable accuracies (Supplementary Fig. 8), combining the estimated features from all layers did not improve visual reconstruction. Multiple reasons are conceivable. Higher layers contain more abstract information and contribute less to the specific pixel patterning (Mahendran and Vedaldi, 2015). The De-CNN reverses the CNN with approximation, especially at the un-pooling step (Zeiler and Fergus, 2014; Springenberg et al., 2014). As a result, the decoding errors cascade down the CNN, causing accumulated errors in the reconstructed pixels.

In this study, the fMRI-decoded visual reconstruction emphasized foreground and suppressed background (Fig. 8). This intriguing finding is likely attributable to the effects from both bottom-up salience (Itti and Koch, 2001) and top-down attention (Desimore and Duncan, 1995). The CNN captures visual salience (Simonyan et al., 2013; Canziani and Culurciello, 2015), but has no mechanism for top-down attention. It thus helps to dissociate the salience vs. attention effects. To explore the effects from salience but not attention, we applied the decoding model to the fMRI signals predicted by the voxel-wise encoding models. As in Supplementary Fig. 9, the resulting visual reconstruction also highlighted the foreground objects. It suggests that visual salience is captured by the CNN and indeed contributes to the foreground selectivity. However, decoding of the measured fMRI signals revealed even more focal emphases on foreground objects (Supplementary Fig. 9). Therefore, in addition to bottom-up salience, there are other selection mechanisms, likely top-down attention (Desimore and Duncan, 1995) that shape the fMRI responses during movie viewing.

**Direct decoding of semantic representations and categorization**

This study also demonstrates the value of using the CNN to directly decode and categorize semantic representations. The CNN contains a semantic space in its $2^{nd}$ highest layer. It supports object recognition in the output layer with either finely or coarsely defined categories, and is even transferrable to other vision tasks (Razavian et al., 2014). Hence, it represents a generalizable semantic space, emerging progressively from the visual features in the lower levels. The decoding model allows us to directly estimate the representation in this semantic space for arbitrary natural stimuli. The decoded semantic representation is generalizable and transferable, and independent of the definition of categories, unlike the categorical decoding method recently reported elsewhere (Huth et al., 2016b).

In addition, the semantic space in the CNN can be readily converted to human-defined categorical labels, by training a classifier to match the semantic representation to the label. It effectively translates a vector representation to a word, and allows the textual interpretation of brain activity. The classifier can be trained without redefining the semantic space, by only retraining the CNN's output layer with labeled images. So, the classifier is separate from the decoding model. This offers interesting extensions of the current decoding capabilities. One may utilize the ever-expanding labeled images to set up various interpretations of the semantic representations decoded from brain activity.

**Methodological considerations**

The bivariate (voxel-to-unit) correlation analysis is a simple way to explore the correspondence between the brain and the CNN. It does not require data-demanding training for any encoding model, and thus applicable to data with limited length. Despite its simplicity, the correlation analysis is effective for mapping multiple organizational patterns in cortical representations during natural vision, including the cortical retinotopy (Fig. 2.b), hierarchy (Fig. 2.c), and category representation (Fig. 2.d and 2.e). As such, this analysis, along with the natural-vision paradigm, is a good strategy for multi-purpose functional mapping, arguably more preferable than conventional localizer or mapping paradigms. However, the bivariate analysis has two major limitations. A one-to-one correspondence between brain voxels and CNN units, is not strictly plausible; the correlation does not account for the computation at a voxel or region. Both limitations are addressable with the CNN-based voxel-wise encoding models, if sufficient training data are available.

The demand for large training data limits the practical utility of the CNN-based encoding models for the individual-subject analysis. It is not realistic to acquire hours of data, just for model training. Thus, it remains challenging to expand the analysis from a few subjects to a large number of subjects, as required for typical imaging studies. One way to address this practical limitation is to train models with sufficient data from one or few subjects, and extrapolate the trained models to other subjects. Results in this study support this feasibility. The encoding and decoding models could be transferred across subjects, yielding reduced yet still significant prediction and decoding accuracy. These results are consistent with previous findings that cortical responses to naturalistic stimuli are highly consistent across subjects (Hasson et al., 2004; Russ and Leopold, 2015; Lu et al., 2016). They also confirm that anatomical registration succeeds at matching up functionally similar areas. However, further improvement in inter-subject encoding and decoding is still desirable. It requires future methodological development to improve the anatomical and/or functional alignment, in order to account for individual differences and support group-level analysis.

When training the encoding and decoding models, regularization is necessary for mitigating model over-fitting. In this study, we prefer L2 to L1 regularization in training the voxel-wise encoding model. L1 regularization was too computationally demanding to be realistic for training and cross-validating an encoding model for each and every voxel in the brain (59,412 voxels), with a large feature space despite dimension reduction. In contrast, we prefer L1 to L2 regularization in training the decoding models, because the former resulted in a better decoding performance than the latter (Supplementary Fig. 12.a), with more yet affordable computational expense.

When training the encoding models, the training efficiency and robustness were improved given the PCA-based dimension reduction. After the dimension reduction, the feature representations effectively redistributed to a similar statistical distribution as the fMRI signal, mitigating the need for log-transformation without compromising the model-prediction performance (Supplementary Fig. 12.b). Furthermore, we used a canonical HRF to account for neurovascular coupling at every voxel. This may or may not be ideal, given the potential variation in HRF across locations. However, a fixed HRF model is a conservative model-fitting strategy, such that the model-prediction performance reflects the appropriateness of the feature models. In an exploratory investigation, we also took the HRF peak latency as a hyper-parameter when training and cross-validating the encoding models. The voxel-wise latency with optimal cross-validation performance was on average around 4s (Supplementary Fig. 12.c), consistent with the latency in the fixed HRF model.

**Future Directions**

The CNN still falls short for modeling and explaining visual-cortical activity during dynamic natural vision (Fig. 4). Future studies should be directed towards models that include not only feedforward, but also feedback (Kafaligonul et al., 2015) and recurrent (Polack and Contreras, 2012) connections. A model that best matches to the brain is expected to reflect the brain's architectures and principles. The human visual cortex may use a deeper hierarchy, than the 8-layer CNN in this study, spanning >20 visual areas (Wandell et al., 2007), and thus it may be better explained by deeper CNNs (Simonyan and Ziserman, 2014; He et al., 2015). Complementary to CNN, recurrent neural networks (Donahue et al., 2015; Srivastava et al., 2015) account for temporal structure in videos, learning spatiotemporal representations more effectively than CNNs that take multiple video frames as input (Tran et al., 2014). Feedback connections may be added to further reflect the brain's predictive coding (Rao and Ballard, 1999), or attention selection (Stollenga et al., 2014). Other plausible models are generative in nature (Dayan et al., 1995; Kingma and Welling, 2013), in line with the free-energy theory – a likely principle of the brain (Friston and Kiebel, 2009; Friston, 2010). Such models are worth exploring, individually or in combination, to better explain brain activity during natural vision. Matching network models to the brain, may lead to better systems for artificial intelligence (AI) (Yamins et al., 2014; Fong et al., 2017).

It is desirable to evaluate and compare different models in explaining the brain's responses to natural visual stimuli, as initially explored elsewhere (Yamins et al., 2014; Khaligh-Razavi and Kriegeskorte, 2014). Objective model comparison requires efforts in making available large open data and open source, along with standardized performance measures, ideally in an open-competition format. Data in different studies are with different types (static vs. dynamic) of stimuli, of different length and quality, with different voxel size and signal to noise ratios (SNR), and from different subjects. Hence, one must be cautious in comparing the quantities of model performance across studies, and carefully exercise statistical significance tests. In particular, model comparison with the correlation-based measures of the prediction accuracy should account for the difference in the number of samples (or the degree of freedom) and the SNR, and be evaluated against the explainable variance in the voxel level (Wu et al., 2006). Per-area summary statistics, although quantitative, may not be ideal, given the variation within each area and the variation of areal definitions between studies and between subjects.

Future studies will benefit from data acquired with more natural images or videos. A more diverse set of natural visual stimuli is expected to further improve the reliability and generalizability of the encoding and decoding models, providing a common source for researchers to evaluate and compare AI models. A general strategy may entail presenting different stimuli to different subjects and then combining the models across subjects, or across labs.

What would also be desirable is the use of neural imaging or recording with higher resolution and sensitivity. For example, the visual reconstruction based on the decoded fMRI activity was blurry and did not contain visual details in texture and color (Fig. 8). This limitation is expected to limit the ability for resolving cluttered scenes. Such information is coded in spatiotemporal activity patterns that are difficult to resolve or distinguish with fMRI at the present resolution. While the decoding models utilized all the voxels in the visual cortex, the visual reconstruction received relatively more contributions from lower visual areas. Voxels in color-specialized areas (e.g. V4) did not lead to more reliable visual reconstruction in this study, whereas another study has shown the initial promise of decoding colors (Hsieh and Tse, 2010).

The deep-learning-enabled brain decoding described here as a means to recreate dynamic visual experience has significant potential for reading and reconstructing other sensory or cognitive experiences as well. Since deep-learning models are already available for speech recognitions (Hinton et al., 2012) and language processing (Collobert and Weston, 2008), decoding of brain measures in response to natural hearing, speech, and language are realistically attainable goals (Huth et al., 2016a). Likewise, since we know that sensory images, memories and dreams involve neural substrates that overlap with those for real sensation (Kosslyn et al.,

1997; Horikawa et al., 2013), it is foreseeable that deep-learning models would also be potentially successful in decoding the internal images of the human mind.

**Supplementary Information**: Supplementary material is available on *Cerebral Cortex*.

https://doi.org/10.1093/cercor/bhx268.

**Open data**: The data and source codes related to this study are online available in https://engineering.purdue.edu/libi/lab/Resource.html.

**Acknowledgements**: This work was supported in part by NIH R01MH104402. The authors thank Eugenio Culurciello and Alfredo Canziani for their technical support in the CNN, to Gregory Francis and Paula Leverage for insightful discussions and suggestions in the paper. The authors also thank the reviewers who provide constructive comments about earlier versions of this paper, prior to its publication.